\documentclass[11pt, a4paper]{article}

\usepackage{indentfirst}
\usepackage{geometry}
\usepackage{threeparttable}
\usepackage{rotating}
\usepackage{setspace}
\usepackage{placeins}
\usepackage[hidelinks]{hyperref}
\usepackage{amsmath}
\usepackage{amssymb}
\usepackage{amsfonts}
\usepackage{graphicx}
\usepackage{enumitem}

\usepackage{dsfont}
\usepackage{subcaption} % Replacing subfigure with subcaption
\usepackage{lscape}
\usepackage{float}
\usepackage[running]{lineno}
\floatstyle{boxed}
\restylefloat{figure}
\usepackage{multirow}
\usepackage[compress]{natbib}
\usepackage{booktabs}

\floatstyle{plain}
\restylefloat{figure}
\usepackage{color}

\usepackage{textcomp}
\usepackage{tabularx}
\usepackage{makecell}
\usepackage{comment}

\hypersetup{
     colorlinks = true,
     linkcolor = black,
     citecolor = blue
}

\usepackage{xcolor}
\usepackage{colortbl}
\usepackage{adjustbox}

\usepackage{caption}
\usepackage{capt-of}
\usepackage{afterpage}

\usepackage{setspace}
\begin{document}

%\linenumbers %  linenumbers on

\title{Unified GARCH-Recurrent Neural Networks in Financial Volatility Forecasting}
\author{Jingyi Wei\footnote{ School of Business, Stevens Institute of Technology,  jwei14@stevens.edu. }, \quad  Steve Yang\footnote{ School of Business, Stevens Institute of Technology,  syang14@stevens.edu. }, \quad and \quad  
Zhenyu Cui\footnote{Corresponding author. School of Business, Stevens Institute of Technology, One Castle Point Terrace, Hoboken, NJ 07030, United States.  zcui6@stevens.edu. Phone: +1(201) 216-5541} 
}

%         \small School of Business\\
%         \small Stevens Institute of Technology\\
%         \small Hoboken, NJ, United States \\
%         \small \{jwei14, syang14, zcui6\}@stevens.edu \\
% }
%\date{July 2024}
\date{}

\maketitle

\begin{abstract}
In this study, we develop a unified volatility modeling framework that embeds GARCH dynamics directly within recurrent neural networks. We propose two interpretable hybrid architectures, GARCH-GRU and GARCH-LSTM, that integrate the GARCH(1,1) volatility update into the multiplicative gating structure of GRU and LSTM cells. This unified design preserves economically meaningful GARCH parameters while enabling the networks to learn nonlinear temporal dependencies in financial time series.
Comprehensive out-of-sample evaluations across major U.S. equity indices show that both models consistently outperform classical GARCH specifications, pipeline-style hybrids, and neural baselines such as the Transformer across multiple metrics (MSE, MAE, SMAPE, and out-of-sample $R^2$). Within this family, the GARCH-GRU achieves the strongest accuracy–efficiency tradeoff, training nearly three times faster than GARCH-LSTM while maintaining comparable or superior forecasting accuracy under normal market conditions and delivering stable and economically plausible parameter estimates.
The advantages persist during extreme market turbulence. In the COVID-19 stress period, both architectures retain superior forecasting accuracy and deliver well-calibrated 99\% Value-at-Risk forecasts, achieving lower violation ratios and competitive Pinball losses relative to all benchmarks.
Overall, the findings underscore the effectiveness of embedding GARCH dynamics within recurrent neural architectures, yielding models that are accurate, efficient, interpretable, and robust for real-world risk-aware volatility forecasting.

\noindent


\end{abstract}

\setlength{\baselineskip}{0.28in}
\renewcommand{\arraystretch}{1.5}

\maketitle

\newpage
\section{Introduction}
Volatility modeling and forecasting have long stood at the forefront of financial econometrics, underpinning crucial tasks such as risk management, derivatives pricing, portfolio optimization, and regulatory compliance. As a fundamental measure of financial market uncertainty and risk, volatility plays a central role in both academic inquiry and industry applications. Accurate volatility prediction is indispensable for understanding market dynamics, pricing assets under uncertainty, and managing financial exposure.

Financial time series exhibit several empirically observed patterns, often referred to as `stylized facts' of volatility. These include volatility clustering, long memory, asymmetric responses to shocks (leverage effects), heavy-tailed return distributions, and mean-reverting behavior. Capturing these features is essential for any model aiming to provide realistic and reliable forecasts. Among these, volatility clustering and persistence are particularly well-documented: periods of high volatility tend to be followed by high volatility, and low-volatility periods likewise tend to persist. These temporal dependencies contradict the assumptions of constant variance and i.i.d. returns commonly assumed  in classical models, motivating the development of more sophisticated volatility modeling approaches. Another key empirical feature is asymmetry, where negative price shocks tend to have a larger impact on future volatility than positive ones of similar magnitude. This leverage effect further challenges symmetric and linear modeling assumptions.

Traditional econometric approaches, particularly the Autoregressive Conditional Hetero\-skedasticity
(ARCH) model introduced by \cite{engle1982autoregressive} and its generalization, the Generalized ARCH (GARCH) model proposed by \cite{bollerslev1986generalized}, laid the foundation for modern volatility modeling. These models, grounded in strong statistical theory, have proven effective in modeling time-varying conditional variance and capturing volatility clustering. Their interpretability, tractability, and widespread applicability have established them as benchmarks in financial econometrics. However, GARCH-family models rely on linear parametric structures and are often limited in their ability to accommodate non-linear dynamics, structural breaks, and extreme events commonly observed in financial data \citep{hansen2005forecast}. Extensions such as asymmetric GARCH variants, long-memory models such as fractionally integrated GARCH (FIGARCH), and more recently, fractional affine multi-component GARCH frameworks, offer improvements but still rely on fixed functional forms. 
%For example, \cite{augustyniak2022long} integrate the Heston-Nandi option pricing framework with FIGARCH dynamics to jointly capture both short- and long-memory components in volatility evolution.
Recent findings like \citep{augustyniak2022long} suggest that volatility exhibits multi-component persistence and structural heterogeneity that standard GARCH specifications cannot fully capture, underscoring the need for more adaptive and expressive modeling approaches.

Meanwhile, machine learning and deep learning architectures have emerged as powerful alternatives for volatility forecasting. Models such as recurrent neural networks (RNNs), long short-term memory (LSTM) networks \citep{hochreiter1997long}, and gated recurrent units (GRUs) \citep{chung2014empirical} can flexibly approximate nonlinear temporal dependencies and adapt to shifting market regimes. Recent empirical evidence further documents that recurrent deep learning architectures can effectively model nonlinear financial time series, including stock-return dynamics and market-level predictive signals \citep{ThomasChristopher2017}. Despite these advances, however, deep learning models often suffer from limited interpretability and lack the theoretical grounding inherent in econometric models. Their internal dynamics do not explicitly encode the `stylized facts' that econometricians have documented over decades. Moreover, LSTM-based volatility models can be computationally heavy and over-parameterized, limiting their deployability in latency-sensitive settings.

Recent efforts have attempted to bridge the gap between statistical rigor and data-driven flexibility by combining GARCH-type models with deep learning architectures. Existing work largely focuses on LSTM-based hybrids that provide improved flexibility but still treat GARCH outputs as exogenous features, rather than embedding volatility dynamics directly into the recurrent architecture. This static or feed-forward integration limits the extent to which neural networks can interact with or adaptively modulate econometric components. A recent advancement by \cite{zhao2024garch} embeds GARCH regression into the LSTM cell, replacing the output gate and enabling deeper integration of volatility signals. However, their formulation uses scalar hidden states and implicitly assumes zero-mean innovations, which restricts the model’s capacity to capture external drivers, multi-dimensional dependence structures, and richer forms of temporal heterogeneity. Beyond structural limitations, LSTM-based hybrids also face practical constraints. The LSTM architecture carries substantial computational overhead due to its intricate gating structure and large parameter count, leading to heavier training costs and slower inference, particularly in latency-sensitive applications such as high-frequency trading. In contrast, the GRU architecture offers a more streamlined gating mechanism with significantly fewer parameters, yielding meaningful improvements in computational efficiency while retaining strong capacity to model nonlinear and temporal dependencies \citep{chung2014empirical}. The computational and architectural advantages make GRU-based hybrids particularly appealing for real-world deployment.

In this study, we introduce a unified GARCH–Recurrent Neural Network (GARCH-RNN) framework that integrates the interpretability and statistical rigor of traditional econometric models with the nonlinear learning capabilities of deep neural networks. The proposed framework embeds GARCH(1,1) dynamics directly into neural network (NN) cells, preserving the gating mechanisms and hidden state transitions intrinsic to the NN structure. This embedded design enables the model to jointly learn key volatility `stylized facts', such as clustering and persistence, alongside complex temporal patterns in an unified training paradigm, where both the GARCH parameters and NN weights are optimized simultaneously. The architecture further supports hierarchical feature fusion through stacked integrated layers, facilitating multi-scale representation learning that captures heterogeneous dynamics across time horizons. Within this general framework, we develop two specific model instantiations based on GRU and LSTM architectures.
Despite their architectural differences, both variants retain explicit GARCH parameters, which remain economically interpretable even after joint optimization with neural network weights.
To evaluate the robustness and generalizability of the proposed framework, we conduct extensive empirical experiments on a diverse set of financial assets, including the S\&P 500 Index, the Dow Jones Industrial Average (DJI), and the NASDAQ Composite Index, under varying market regimes and volatility conditions. The results demonstrate that the GARCH-RNN framework substantially improves volatility forecasting accuracy relative to both classical GARCH models and state-of-the-art deep learning baselines. Moreover, the complementary behavior of the GARCH-GRU and GARCH-LSTM variants provides insight into the regime-dependent nature of volatility dynamics and highlights the value of integrating econometric structure with neural sequence models.

%This study addresses the aforementioned research gaps by introducing a novel integrated GARCH-GRU framework that unifies the interpretability and statistical rigor of traditional econometric models with the nonlinear learning capabilities of deep neural networks. Specifically, we design an architecture that embeds GARCH(1,1) dynamics directly into GRU cells, preserving the gating mechanisms and hidden state transitions intrinsic to the GRU structure. This embedded design enables the model to jointly learn key volatility `stylized facts', such as clustering and persistence, alongside complex temporal patterns in an end-to-end training paradigm, where both the GARCH parameters and GRU weights are optimized simultaneously. The architecture further supports hierarchical feature fusion through stacked integrated layers, facilitating multi-scale representation learning that captures heterogeneous dynamics across time horizons. To evaluate the robustness and generalizability of the proposed framework, we conduct extensive empirical experiments on a diverse set of financial assets, including the S\&P 500 Index, the Dow Jones Industrial Average (DJI), and the NASDAQ Composite Index, under varying market regimes and volatility conditions. 

The rest of the paper is structured as follows. Section 2 reviews related literature spanning econometric volatility models, deep learning approaches, and recent hybrid methods. Section 3 presents the methodological development of the GARCH-RNN framework, including the preliminaries and the theoretical formulation of the GRU- and LSTM-based variants. Section 4 describes the empirical design, datasets, evaluation metrics, and benchmark models, followed by a comprehensive presentation of the evaluation results. Section 5 concludes with a discussion of the main findings, practical implications, limitations, and avenues for future research.

\section{Related Works}

Classical econometric models, such as the ARCH and GARCH models,  
%Autoregressive Conditional Heteroskedasticity (ARCH) model introduced by \cite{engle1982autoregressive} and its generalization, the Generalized Autoregressive Conditional Heteroskedasticity (GARCH) model by \cite{bollerslev1986generalized}, 
laid the groundwork for capturing time-varying volatility and volatility clustering in asset returns. These models are grounded in a well-defined probabilistic framework and have served as the benchmark in both academic and applied settings.
To accommodate empirical regularities beyond volatility clustering, numerous extensions to the basic GARCH framework have been introduced. The Integrated GARCH (IGARCH) model by \cite{engle1986modelling} incorporates a unit root in the conditional variance process, allowing for shocks to have persistent effects on volatility. The GARCH-in-Mean (GARCH-M) model proposed by \cite{engle1987estimating} explicitly incorporates conditional variance into the mean equation, capturing the risk-return trade-off. To address the observed asymmetry in volatility responses, where negative returns often lead to larger increases in volatility than positive ones, \cite{nelson1991conditional} introduced the Exponential GARCH (EGARCH) model, which models the logarithm of the variance and naturally accommodates asymmetric effects. Similarly, the GJR-GARCH model by \cite{glosten1993relation} includes an indicator function to differentiate the impact of positive and negative shocks, while the Threshold GARCH (TGARCH) model developed by \cite{zakoian1994threshold} models the conditional standard deviation and also captures leverage effects. 
Recognizing the limitations of univariate models in capturing interdependencies across multiple financial time series, multivariate GARCH-type models have also been developed. For instance, the bivariate ARCH model introduced by \cite{Engle1984} represents an early attempt to model co-movements in volatility. The Baba-Engle-Kraft-Kroner (BEKK) model formulated by \cite{engle1995multivariate} provides a flexible yet computationally tractable framework for multivariate volatility dynamics that has been influential in empirical finance application, such as \cite{bollerslev2008glossary}. Subsequent extensions further enhance the ability to model dynamic cross-asset interactions. The Varying Correlation MGARCH model of \cite{Tse2002} captures time‐varying conditional correlations through a parsimonious updating mechanism, while regime-switching MGARCH frameworks such as \cite{So2009} incorporate nonlinear transitions across volatility regimes, thereby accommodating the extreme events, structural breaks, and tail-risk behavior widely documented in empirical studies (e.g., \cite{poon2003forecasting}; \cite{andersen2003modeling}).

While these models offer a robust and interpretable framework, their inherent reliance on linear dynamics and predefined functional forms limits their capacity to capture complex nonlinear relationships, structural breaks, and regime shifts commonly observed in financial markets. Moreover, their parametric nature poses challenges in adapting to rapidly evolving market conditions, particularly under high-frequency and large-scale data environments.

Beyond volatility modeling itself, machine learning methods have shown strong predictive performance across a wide range of financial forecasting tasks. These include equity return prediction using ensemble learners and deep neural networks \citep{krauss2017deep}, as well as large-scale risk forecasting problems characterized by nonlinear interactions and high-dimensional covariates \citep{huck2019large}. Hybrid neural-network architectures have also been explored in financial contexts. For example, \citet{sermpinis2013forecasting} develop an adaptive RBF–PSO model that outperforms ARMA and several neural-network benchmarks in exchange-rate forecasting, illustrating how nonlinear and optimization-enhanced architectures can extract complex temporal patterns from financial time series. Building on these advances, deep learning approaches have been increasingly applied to volatility modeling and forecasting. Recurrent neural networks (RNNs), especially Long Short-Term Memory (LSTM) networks and Gated Recurrent Units (GRUs), are well suited for capturing long-range dependencies and nonlinear dynamics in time series data. The LSTM model, introduced by \cite{hochreiter1997long} and further extended by \cite{Gers2000} and \cite{ThomasChristopher2017}, has shown empirical superiority in capturing sequential patterns in financial datasets. In a comparative study, \cite{liu2019novel} found that LSTM-based models outperform traditional GARCH variants for volatility prediction over longer forecast horizons. GRU, empirically validated by \cite{chung2014empirical}, offers a more computationally efficient alternative to LSTM while maintaining comparable performance in sequence modeling tasks. It has been successfully employed in various financial prediction tasks, including the Two-Stream GRU (TGRU) model proposed by \cite{8456512}, which integrates bi-directional recurrent mechanisms and achieves improved predictive accuracy over classical models.

Despite their remarkable predictive power, deep neural networks often lack financial interpretability and theoretical grounding. Similar concerns have been noted in broader areas of financial risk modeling, where deep learning models, such as those used in bankruptcy prediction, also face persistent challenges related to transparency and explainability \citep{mai2019deep}. The inherent `black-box' nature of these architectures poses challenges for transparency, risk attribution, and regulatory acceptability. Furthermore, they do not explicitly model conditional heteroskedasticity, and thus may overlook important volatility-specific features that are well captured by econometric models.

To reconcile the statistical interpretability of GARCH with the representational flexibility of deep learning, a growing body of literature has explored hybrid approaches. Early efforts typically relied on sequential or decoupled modeling pipelines, where GARCH models were used to extract residuals or conditional variances, which were then passed as inputs into neural networks for downstream prediction. Notable examples of such decoupled architectures include the hybrid models developed by \cite{roh2007forecasting}, \cite{tseng2008artificial}, \cite{hajizadeh2012hybrid}, and \cite{kristjanpoller2014volatility, kristjanpoller2016forecasting, kristjanpoller2017volatility}, as well as \cite{kim2018forecasting}. While effective in improving forecasting accuracy, these approaches typically involve separate training processes for the statistical and neural components, limiting their capacity to learn joint representations and often resulting in inefficiencies and suboptimal integration. Recent studies have proposed end-to-end hybrid architectures that attempt to jointly optimize econometric and deep learning components. For instance, \cite{zhao2024garch} introduced an embedded GARCH-LSTM model that integrates the GARCH regression mechanism within LSTM cells by replacing the output gate. This approach represents a significant step towards unified modeling, enabling direct interaction between GARCH dynamics and temporal feature extraction. However, this model imposes strong assumptions, such as scalar hidden state representations and zero-mean innovations, which restrict its expressiveness and ability to model complex market behaviors and external covariate influences. 

While substantial progress has been made in both traditional volatility modeling and machine learning-based forecasting, existing approaches either lack the interpretability required for financial applications or fail to fully utilize the learning capacity of neural networks. The current literature also tends to emphasize LSTM-based hybrid models, leaving the advantages of GRU-based integrations largely untapped. This motivates the development of new unified frameworks that can simultaneously capture volatility stylized facts and complex temporal dependencies in a computationally efficient and theoretically coherent manner.

\section{Methodology}

\subsection{Theoretical Foundations}
\subsubsection{GARCH Model}
The GARCH model \citep{bollerslev1986generalized} includes lagged conditional variances along with lagged squared residuals, effectively capturing volatility clustering and persistence in financial time series. Assuming a constant mean return and given the information set $\psi_{t-1}$, the model is specified as:

\begin{align}
    &r_t = \mu + \epsilon_t, 
    \label{mean function}\\
    &\epsilon_t|\psi_{t-1} \sim \mathcal{D}(0, \sigma_t^2), \\
    &\sigma_t^2 = \omega_0 + \sum_{i=1}^{p}\alpha_i\epsilon_{t-i}^{2} + \sum_{j=1}^{q}\beta_j \sigma^2_{t-j},
    \label{GARCH}
\end{align}
where $\mathcal{D}(0, \sigma_t^2)$ typically denotes a normal or Student’s t-distribution. The model parameters satisfy $\omega > 0$, $\alpha_i \geq 0$, $\beta_j \geq 0$, and $\sum_{i=1}^p\alpha_i + \sum_{j=1}^q \beta_j < 1$ to ensure positivity and weak stationarity of the conditional variance process. The  coefficients $\alpha_i$ capture the short-term impact of shocks, while the  coefficients $\beta_j$ reflect the persistence of volatility over time. 

\subsubsection{Recurrent Neural Networks}
Both LSTM and GRU models are widely adopted recurrent neural network (RNN) architectures for sequential modeling and have demonstrated strong empirical performance in financial forecasting tasks. These models are particularly effective at capturing temporal dependencies in time series data due to their gating mechanisms, which regulate the flow of information through the network over time.
%LSTM
LSTM is designed to handle long-range dependencies by introducing memory cells and three gating components: input, forget, and output gates. These gates enable the model to selectively retain or discard information from previous time steps, thereby mitigating the vanishing gradient problem associated with standard RNNs. The LSTM cell computes the hidden state and cell state at time $t$ through the following set of equations: 
\begin{align}
f_t &=\sigma\left(W_f x_t + U_f h_{t-1}+b_f\right) 
\label{LSTM equation 1}\\
i_t &=\sigma\left(W_i x_t + U_i h_{t-1}+b_i\right) 
\label{LSTM equation 2}\\
o_t &=\sigma\left(W_o x_t + U_o h_{t-1}+b_o\right)
\label{LSTM equation 3}\\
c_t &=f_t \odot c_{t-1}+i_t \odot \tilde{c}_t
\label{LSTM equation 4}\\
\tilde{c}_t &=\tanh \left(W_c x_t + U_c h_{t-1} + b_c\right)
\label{LSTM equation 5}\\
h_t &=o_t \odot \tanh \left(c_t\right) 
\label{LSTM equation 6}
\end{align}
where $x_t$ is the input vector at time $t$, $h_{t-1}$ is the previous hidden state, $c_{t-1}$ is the previous cell state, and $\sigma(\cdot)$ denotes the sigmoid activation function. The memory cell $c_t$ acts as an internal state that accumulates information over time, while $h_t$ serves as the output at time $t$. The element-wise product operation (Hadamard product) is denoted by $\odot$.

% GRU
The GRU is a streamlined variant of the LSTM that retains the key advantages of gated memory while reducing computational complexity. It employs only two gates, the update gate and reset gate, thereby simplifying the architecture and accelerating training. The GRU computes its hidden state via the following equations:
\begin{align}
z_t & =\sigma\left(W_z x_t + U_z h_{t-1}+ b_z\right) 
\label{GRU equation 1}\\
r_t & =\sigma\left(W_r x_t + U_r h_{t-1}+ b_r\right) 
\label{GRU equation 2}\\
\tilde{h}_t & =\tanh \left(W_h x_t + U_h(r_t \odot h_{t-1}) + b_h\right) 
\label{GRU equation 3}\\
h_t & =\left(1-z_t\right) \odot h_{t-1}+z_t \odot \tilde{h}_t
\label{GRU equation 4}
\end{align}
where $z_t$ and $r_t$ control how much past information is carried forward and how much is reset at each time step, respectively. Unlike LSTM, GRU does not maintain a separate cell state, but instead uses the hidden state $h_t$ for both memory and output.

\subsection{Unified GARCH-RNN Framework}
\subsubsection{GARCH-GRU Model}
\begin{figure}[h!]
\centering
\includegraphics[width = 0.8\textwidth]{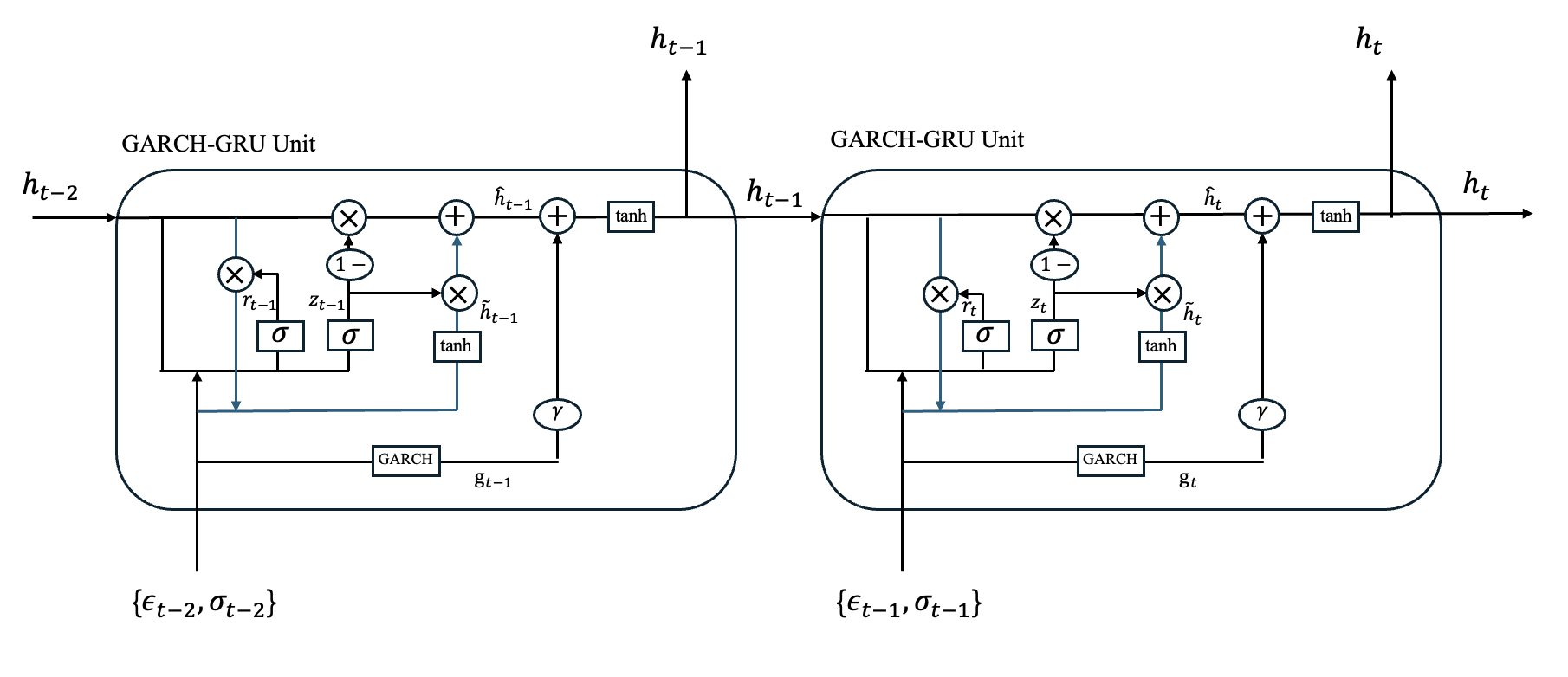} \\
\caption{Graphical Representation of integrated GARCH-GRU}
\label{GARCH-GRU unit}
\end{figure}

The proposed GARCH-GRU model builds upon the conventional GARCH(1,1) specification with constant mean assumption in Eq.\eqref{mean function}, which is formulated as: 
\begin{equation}
    \sigma_t^2 = \omega_0 + \alpha\epsilon_{t-1}^2 + \beta \sigma_{t-1}^2.
    \label{GARCH(1,1)}
\end{equation}
Asset returns $r_t$ exhibit innovations $\epsilon_t$ with time-varying conditional variance $\sigma_t^2$. The key innovation lies in structural integration of GARCH dynamics into GRU operations through a multi-stage gating mechanism, as formalized in Eqs.\eqref{GARCH-GRU 0}-\eqref{GARCH-GRU 5}.
\begin{align}
g_t &= \phi\left(\omega_0+\alpha \epsilon_{t-1}^2+\beta \sigma_{t-1}^2\right)
\label{GARCH-GRU 0} \\
z_t&=\sigma\left(W_z x_t+U_z h_{t-1}+b_z\right) 
\label{GARCH-GRU 1}\\
r_t&=\sigma\left(W_r x_t+U_r h_{t-1}+b_r\right) 
\label{GARCH-GRU 2}\\
\tilde{h}_t&=\tanh \left(W_h x_t+U_h\left(r_t \odot h_{t-1}\right)+b_h\right) 
\label{GARCH-GRU 3}\\
\hat{h}_t&=\left(1-z_t\right) \odot \tilde{h}_t +z_t \odot h_{t-1}
\label{GARCH-GRU 4}\\
h_t&=\tanh\left(\hat{h}_t+\gamma g_t\right)
\label{GARCH-GRU 5}
\end{align}
The GARCH-GRU cell operates through dual processing streams. First, the GARCH component generates volatility estimates $g_t$ using recombination of past residuals and variances (Eq.\ref{GARCH-GRU 0}), where $\phi(u) = W_g u + b_g$ denotes a linear projection aligning scalar GARCH output with GRU hidden state dimensions. Second, a modified GRU architecture processes sequential inputs through standard reset/update gates (Eqs.\eqref{GARCH-GRU 1}-\eqref{GARCH-GRU 4}), followed by additive integration of GARCH signals into the hidden state update (Eq.\eqref{GARCH-GRU 5}). The learnable parameter $\gamma$ governs the contribution intensity of GARCH components, while element-wise addition enables granular modulation of individual hidden units. The formulation in Eq.\eqref{GARCH-GRU 5} can be equivalently expressed in its inverse tanh form as
\begin{equation}
    \frac{1}{\gamma} \tanh ^{-1}\left(h_t\right)=g_t+\frac{1}{\gamma} \hat{h}_t
    \label{interpretability}
\end{equation}
This transformation reveals that the GARCH output $g_t$ and the recurrent component $\hat{h}_t$ contribute additively to the latent pre-activation of the hidden state. In this unsaturated representation, $g_t$ acts as a volatility-driven bias shift, while $\hat{h}_t$ captures feature-dependent temporal dynamics. The parameter $\gamma$ regulates their relative influence: larger values amplify the volatility contribution, whereas smaller values recover standard GRU dynamics. Because $\tanh ^{-1}(\cdot)$ is monotone, this decomposition provides an interpretable mapping between econometric variance processes and neural hidden representations, clarifying how conditional volatility information modulates the recurrent state evolution.
This design preserves GRU's capacity to model nonlinear temporal dependencies while explicitly encoding volatility persistence patterns. A graphical representation of the integrated GARCH-GRU framework is shown in Figure.\ref{GARCH-GRU unit}.

To ensure stability of the GARCH component, we apply a reparameterization to enforce the stationarity constraints on parameters ($\omega_0$, $\alpha$, $\beta$), especially for  $\alpha + \beta < 1$. The GRU subsystem employs dropout regularization to prevent overfitting while maintaining gradient flow through deep layers. Final volatility predictions are generated through a constrained output layer that applies softplus activation followed by square-root transformation, ensuring non-negative volatility estimates consistent with financial time series properties. 

This integrated framework achieves three complementary objectives: explicit representation of volatility clustering through GARCH components, adaptive extraction of complex temporal dependencies via GRU networks, and preservation of econometric validity through interpretable parameter constraints. 
The additive integration mechanism formalized in Eq.\eqref{GARCH-GRU 5} establishes a mathematically grounded interface between autoregressive variance dynamics and learned neural representations. This synergistic combination enables joint optimization of statistical and deep learning components while retaining structural interpretability, ensuring that volatility forecasts simultaneously capture conventional persistence patterns and nonlinear feature interactions inherent in financial time series.

\subsubsection{GARCH-LSTM Model}
To facilitate a systematic comparison between recurrent architectures in hybrid volatility modeling, we construct a GARCH-LSTM model that parallels the design of the proposed GARCH-GRU framework, as formalized in Eqs.\eqref{GARCH-LSTM 0}–\eqref{GARCH-LSTM 7}. Consistent with the GARCH-GRU’s integration strategy, the GARCH-LSTM model introduces a volatility-driven multiplicative gate derived from the standard GARCH(1,1) specification (Eq.\eqref{GARCH-LSTM 0}), which encodes key stylized facts of volatility, including clustering and persistence.

While both architectures incorporate GARCH signals to modulate neural dynamics, they diverge in the final fusion mechanism. In the GARCH-GRU model, the GARCH component is \textit{additively} combined with the intermediate hidden state (Eq.\eqref{GARCH-GRU 5}), allowing for element-wise modulation controlled by a learnable scaling parameter. In contrast, the GARCH-LSTM model adopts a \textit{multiplicative} modulation strategy, partially inspired by the integration scheme proposed in \cite{zhao2024garch}. Specifically, the final hidden state $h_t$ is obtained by scaling the LSTM’s intermediate output $\tilde{h}_t$ with a learnable coupling parameter $w$ and a nonlinear transformation of the GARCH component $g_t$ (Eq.\eqref{GARCH-LSTM 7}). This multiplicative gating mechanism  allows  $w$ to adaptively calibrates the magnitude of GARCH contributions relative to the LSTM’s intrinsic temporal dynamics. As $w$ increases, the magnitude of $g_t$ becomes larger, thereby intensifying its modulation effect on the final hidden state and increasing its contribution to the model’s volatility forecast.
\begin{align}
g_t &= \phi\left(\omega_0+\alpha \epsilon_{t-1}^2+\beta \sigma_{t-1}^2\right)
\label{GARCH-LSTM 0} \\
f_t&=\sigma\left(W_f x_t + U_f h_{t-1} +b_f\right) 
\label{GARCH-LSTM 1}\\
i_t&=\sigma\left(W_i x_t +U_i h_{t-1} +b_i\right) 
\label{GARCH-LSTM 2}\\
o_t&=\sigma\left(W_o x_t + U_o h_{t-1} +b_o\right) 
\label{GARCH-LSTM 3}\\
\tilde{c}_t&=\tanh \left(W_c x_t +U_c h_{t-1} +b_c\right) 
\label{GARCH-LSTM 4}\\
c_t&=f_t \odot c_{t-1}+i_t \odot \tilde{c}_t
\label{GARCH-LSTM 5}\\
\tilde{h}_t &= o_t \cdot tanh(c_t) 
\label{GARCH-LSTM 6}\\
h_t&=\left(1+ w \cdot \tanh \left(g_t\right)\right) \odot \tilde{h}_t
\label{GARCH-LSTM 7}
\end{align}
Unlike the GARCH-GRU’s element-wise additive fusion, which locally adjusts individual hidden units, the GARCH-LSTM’s approach globally amplifies or attenuates the entire hidden state proportionally to the volatility estimate $g_t$. And the $tanh(\cdot)$ normalization ensures bounded scaling while preserving the directional consistency of hidden state updates, thereby enabling the model to emphasize high-volatility regimes without destabilizing gradient propagation.

\section{Empirical Study}

\subsection{Experimental Design}
\textbf{Datasets}: We collect daily closing prices $p_t$ of three major U.S. stock indices, S\&P 500, Dow Jones Industrial Average (DJI), and NASDAQ Composite, over the period from January 1, 2010, to December 31, 2020. The sample covers multiple volatility regimes, from prolonged low-volatility intervals to sharp turbulence, including the 2020 COVID-19 crisis, providing a comprehensive testing ground for model evaluation. To construct the return series, we follow standard practices in the financial econometrics literature \citep{roh2007forecasting,bucci2017forecasting,kim2018forecasting,bucci2020realized} and compute daily log returns as:

\begin{equation}
    r_t := log{\frac{p_t}{p_{t-1}}}.
    \label{return}
\end{equation}
The realized volatility used as the target variable for forecasting is computed using a rolling window of $k = 5$ days (approximately one trading week) as follows:

\begin{equation}
\begin{aligned}
    \sigma_t &:= \sqrt{\frac{1}{k}\sum_{i=1}^{k}(r_{t-i} - \mu_k)^2} \\
    \mu_k &:= \frac{1}{k}\sum_{i=1}^kr_{t-i}
    \label{volatility}
\end{aligned}
\end{equation}
The sample is divided chronologically to ensure strict out-of-sample evaluation. Observations prior to 2019 are used for model development, where rolling input windows are constructed from this pre-2019 segment to generate training sequences. And the last 20\% of this pre-2019 data is reserved for validation during hyperparameter tuning. Out-of-sample evaluation is conducted primarily on the 2019 period, which represents a relatively stable, low-volatility market regime. To assess robustness in stressed environments, we additionally perform a complementary evaluation on the 2020 sample. The extended forecasts are generated using the same parameters calibrated from the pre-2019 setup, ensuring a genuine out-of-sample stress test reflective of the COVID-19 volatility shock.

\begin{table}[!htbp]
\centering
\caption{Summary Statistics of the Datasets}
\label{tab:DATASET statistics}
\renewcommand{\arraystretch}{1.25}
\resizebox{0.95\textwidth}{!}{
\begin{tabular}{lcccccccc}
\toprule
\textbf{Dataset} & \textbf{Count} & \textbf{Mean} & \textbf{Std. Dev.} & 
\textbf{Skewness} & \textbf{Kurtosis} & \textbf{ADF} & 
\textbf{Non-zero Mean t-test} & \textbf{LM} \\
\midrule
S\&P 500 & 2658 & 0.0422 & 1.1108 & $-0.9122$ & 16.6378 & 
$-12.5060^{***}$ & $1.9627^{**}$ & $1042.7836^{***}$ \\
DJI & 2658 & 0.0382 & 1.1056 & $-1.0128$ & 22.3001 & 
$-12.8978^{***}$ & $1.7794^{*}$ & $1059.0030^{***}$ \\
NASDAQ & 2658 & 0.0587 & 1.2327 & $-0.8274$ & 11.1028 & 
$-14.5179^{***}$ & $2.4537^{**}$ & $887.2649^{***}$ \\
\bottomrule
\end{tabular}
}
%\vspace{2mm}
\begin{tablenotes}
\scriptsize
\item `*' indicates significance at the 10\% level, `**' at the 5\% level, and `***' at the 1\% level. `ADF' is the Augmented Dickey-Fuller test for unit-root existence. `LM' is the ARCH Lagrange Multiplier test for heteroscedasticity.
\end{tablenotes}
\end{table}

Table.\ref{tab:DATASET statistics} presents summary statistics and diagnostic tests for the return series of each index. All return series reject the null hypothesis of the Augmented Dickey-Fuller (ADF) test at the 1\% significance level, confirming that the return series are stationary and do not contain unit roots. Additionally, the Lagrange Multiplier (LM) test for ARCH effects strongly rejects the null hypothesis of homoscedasticity for all three indices, indicating the presence of conditional heteroskedasticity, and thereby justifying the application of GARCH-type models. Furthermore, a two-sided t-test is conducted to assess whether the mean return is statistically different from zero. The resulting t statistics reject the null hypothesis of zero mean for all indices, supporting the constant but non-zero mean specification adopted in our volatility models for the subsequent empirical analysis.

\begin{figure}[h!]
\centering
\includegraphics[width = 0.8\textwidth]{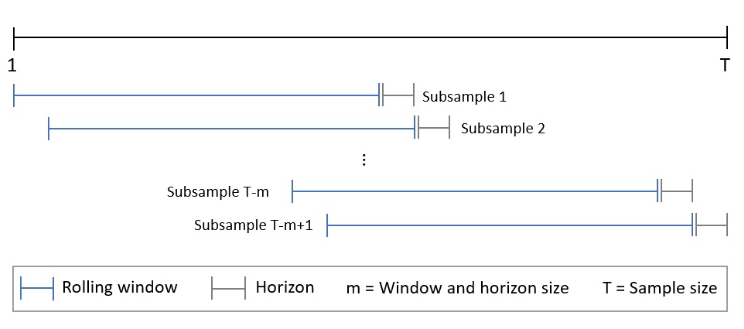} \\
\caption{Rolling Window Methodology}
\label{rolling window}
\end{figure}

All models are trained and evaluated using a rolling window approach (see Fig.\ref{rolling window}) to simulate realistic forecasting conditions. A fixed-length window of 22 trading days (approximately one month) is used to train the model, which is then used to generate volatility forecasts for the next period. The forecast horizons considered in this study are 1-day, 3-day, and 7-day (1-week) ahead forecasts. After each forecast, the window is rolled forward by one day, and the process is repeated until the entire test set is covered. This strategy ensures consistency in evaluation and reflects the operational setting in real-time forecasting scenarios.

\textbf{Model Configuration}: The proposed GARCH–GRU and GARCH–LSTM architectures are constructed by introducing custom modifications to the standard GRU and LSTM cells, embedding conditional heteroskedasticity dynamics directly within their gating mechanisms in a manner consistent with the GARCH(1,1) specification. This integrated design allows each model to jointly learn nonlinear temporal dependencies and time-varying volatility structures in an end-to-end framework, thereby capturing both econometric stylized facts and complex recurrent dynamics inherent in financial time series.
Model hyperparameters are tuned using the Optuna optimization framework, which employs a Bayesian search strategy to efficiently explore the parameter space. To mitigate overfitting, we apply early stopping based on validation performance, using a patience threshold of 20 epochs with mean squared error (MSE) as the monitoring criterion. Model training is performed using the Adam optimizer, providing adaptive learning rates and stable convergence properties. In addition, a dynamic learning-rate scheduler is employed, reducing the learning rate by a factor of two when validation performance stagnates. This training configuration ensures numerical stability and strong generalization across forecasting horizons, supporting the empirical robustness of the proposed GARCH-RNN hybrid volatility models.

\textbf{Evaluation Framework}:We assess the forecasting performance of the proposed models using a set of complementary statistical accuracy measures, including Mean Squared Error (MSE), Mean Absolute Error (MAE), out-of-sample 
$R^2$ and the Symmetric Mean Absolute Percentage Error (SMAPE). These metrics quantify the deviation between predicted volatility ($\hat{\sigma}_t$) from realized volatility ($\sigma_t$) over the out-of-sample horizon. MSE penalizes large errors more heavily and is therefore sensitive to extreme deviations, while MAE provides a more robust measure of average error magnitude. The out-of-sample $R^2$ evaluates improvement relative to a mean benchmark, offering a scale-adjusted indicator of predictive power. SMAPE captures proportional forecast errors, facilitating comparisons across volatility regimes that differ in scale.

%The Mean Squared Error is defined as
%$$
%\mathrm{MSE}:=\frac{1}{n} \sum_{t=1}^n\left(\hat{\sigma}_t-\sigma_t\right)^2,
%$$
%where $\hat{\sigma}_t$ and $\sigma_t$ denote the predicted and realized volatility at time $t$, respectively. MSE penalizes larger forecast errors more heavily due to the form of the quadratic loss function, making it particularly sensitive to extreme deviations.
%In contrast, the Mean Absolute Error is calculated as
%$$
%\mathrm{MAE}:=\frac{1}{n} \sum_{t=1}^n\left|\hat{\sigma}_t-\sigma_t\right|,
%$$
%which provides a more \textit{robust} measure of average forecast accuracy that is less influenced by outliers. Together, MSE and MAE offer a comprehensive view of both the magnitude and variability of forecast errors, facilitating comparative evaluation across benchmark models and robustness checks.

To complement the statistical accuracy metrics, we evaluate the practical relevance of the volatility forecasts through Value-at-Risk (VaR) backtesting. In addition to examining VaR violations and violation ratios, we compute the pinball loss, which directly assesses the quality of quantile (VaR) forecasts by penalizing asymmetric under- and over-prediction. Incorporating the pinball loss provides a more sensitive assessment of tail-risk forecasting performance and strengthens the evaluation of the models’ utility in risk management applications.

\textbf{Benchmarks}: To rigorously evaluate the forecasting performance of the proposed integrated GARCH-RNN framework, and in particular its GRU- and LSTM-based instantiations, we benchmark the models against several representative classes of approaches that span the main paradigms in the volatility forecasting literature. This set of benchmarks includes traditional econometric models, standard deep learning architectures, and alternative hybrid designs, thereby providing a comprehensive and interpretable basis for performance comparison.
The first benchmark class consists of two-stage rolling hybrid models, where the GARCH and neural components are estimated sequentially rather than jointly. In this framework, denoted `bm\_GARCH\_LSTM', the GARCH(1,1) model is used to generate adaptive volatility features, specifically $\alpha_t \varepsilon_t$ and $\beta_t \sigma_t$. Then these features are used as inputs to a LSTM network to forecast realized volatility over multiple horizons. This decoupled, pipeline-style integration enables the GARCH component to capture short-term volatility clustering and persistence, while the LSTM learns higher-order nonlinear temporal dependencies. The Two-Stage `bm\_GARCH\_LSTM' thus serves as a representative and widely adopted hybrid benchmark, highlighting the difference between traditional pipeline integration and the fully embedded GARCH-RNN architectures proposed in this study.
The second class consists of pure deep learning models, with particular emphasis on the Transformer architectures. These networks are capable of learning rich temporal patterns and nonlinear dependencies, but lack explicit econometric mechanisms for conditional heteroskedasticity. Their inclusion allows us to assess the incremental predictive value gained by embedding structured volatility dynamics directly within recurrent neural networks.
The last class includes traditional econometric models, specifically GARCH(1,1) and GJR-GARCH(1,1,1). These models are well-established in the literature for capturing conditional heteroskedasticity and leverage effects in financial return series. Their inclusion provides a point of comparison for assessing the performance gains achieved through hybridization with neural networks.
Together, these benchmark classes, comprising classical models, standard deep learning architectures, and pipeline-based hybrid designs, form a comprehensive evaluation framework. This allows us to position the proposed GARCH-RNN model within a broad methodological context, ensuring that observed performance improvements are both robust and interpretable, and reflect meaningful advances over existing approaches in the literature.

%\textbf{Architectural Validation}:
Prior to the construction of the integrated GARCH-GRU/LSTM architecture, we conducted a rigorous validation of the custom-implemented neural network architectures to ensure their correctness, reliability and equivalence to the standard Pytorch counterparts. 
The objective of these tests was to confirm that, when operating in a pure recurrent mode without any GARCH-specific components, the custom models yield outputs and gradients that are numerically identical to those produced by the canonical PyTorch implementations under the same initialization and inputs.

Both forward and backward equivalence checks were performed to establish full consistency. In the forward pass, each custom model and its PyTorch reference were provided with identical random input sequences and initial hidden states. The resulting outputs, including the complete hidden-state sequences and the final hidden (and cell) states, were compared under strict numerical tolerances, requiring relative and absolute errors below $10^{-6}$ to guarantee equality up to machine precision. In the backward pass, identical scalar losses were backpropagated through both networks, and the resulting gradients with respect to inputs, hidden states, and model parameters were compared element-wise, confirming that the gradient propagation was exactly matched. These tests were repeated under multiple random seeds to ensure reproducibility and robustness.
All equivalence experiments passed within the prescribed numerical tolerances, demonstrating that our custom recurrent architectures (GRU and LSTM) are mathematically and computationally indistinguishable from the corresponding PyTorch implementations. This validation provides a solid foundation for the subsequent development of the GARCH-GRU and GARCH-LSTM hybrid models, ensuring that any observed differences in empirical performance can be attributed solely to the introduction of the GARCH mechanism rather than to inconsistencies in the underlying recurrent dynamics.

\subsection{Model Evaluation}
\subsubsection{Normal-Period Evaluation (2019 Out-of-Sample)}
Having established the experimental setup, we first examine the empirical forecasting performance of the proposed GARCH-RNN models during the 2019 out-of-sample period. Tables.\ref{tab:metric_mse}–\ref{tab:metric_r2} report the mean accuracy metrics across 20 independent runs for all models, with deterministic GARCH-type benchmarks shown without standard deviations. By comparing the proposed integrated GARCH-GRU and GARCH-LSTM structures against classical econometric models, standard deep learning architectures, and pipeline-style hybrids, the evaluation provides a rigorous and comprehensive assessment of forecasting capability.

Across all three indices, the results consistently show that the GARCH-RNN (both GARCH-GRU and GARCH-LSTM) framework outperforms all benchmark models. This is most clearly reflected in the mean squared error (MSE) shown in Table.\ref{tab:metric_mse}, which places greater weight on large deviations and therefore serves as a stringent measure of forecasting precision. 
The GARCH-GRU attains the lowest MSEs for the S\&P 500 at all forecasting horizons, illustrating its strong ability to model short-term volatility shocks while preserving the persistence characteristic of financial volatility. Its advantage becomes even more pronounced in the NASDAQ dataset at medium and long horizons (3- and 7-day), where the GARCH-GRU clearly dominates. The Transformer performs competitively at very short horizons, but its accuracy deteriorates more rapidly as the horizon lengthens, whereas the GARCH-GRU maintains clear stability.
For the Dow Jones (DJI) dataset, the GARCH-LSTM delivers a slightly lower MSE at the 1-day and 3-day horizon, indicating its strength in capturing highly localized short-term patterns. However, this advantage disappears as the forecast window extends. At the 7-day horizon, although the Transformer performs slightly better, both GARCH-GRU and GARCH-LSTM remain competitive, with the GARCH-GRU outperforming the GARCH-LSTM. These results suggest that although the LSTM-based variant can excel in very short-horizon settings for certain markets, the GRU-based architecture offers stronger robustness, especially when multi-step temporal dynamics are involved. The pipeline hybrid (bm\_GARCH\_LSTM), which relies on sequential rather than joint learning, performs markedly worse, confirming that the performance gains arise from embedding GARCH dynamics directly into the recurrent architecture rather than merely appending volatility features to a standard neural network. Traditional GARCH and GJR-GARCH models yield substantially higher errors across all datasets, demonstrating the limitations of fixed parametric forms in capturing the nonlinear and regime-dependent behavior of realized volatility.

% ===================== MSE =====================
\begin{table}[htbp]
\centering
\caption{Forecasting Performance in Normal-Period (Metric: MSE)}
\label{tab:metric_mse}
\setlength{\tabcolsep}{5pt}
\resizebox{\textwidth}{!}{
\begin{tabular}{llcccccc}
\toprule
\textbf{Dataset} & \textbf{Horizon} &
\textbf{GARCH\_GRU} & \textbf{GARCH\_LSTM} &
\textbf{bm\_GARCH\_LSTM} & \textbf{Transformer} &
\textbf{GARCH(1,1)} & \textbf{GJR-GARCH} \\
\midrule
\multirow{3}{*}{\textbf{S\&P 500}}
& 1 & \makecell{\textbf{0.0183}\\(0.0009)} & \makecell{0.0223\\(0.0020)} & \makecell{0.0539\\(0.0013)} & \makecell{0.0191\\(0.0012)} & 0.0795 & 0.0863 \\
& 3 & \makecell{\textbf{0.0360}\\(0.0022)} & \makecell{0.0374\\(0.0026)} & \makecell{0.0720\\(0.0019)} & \makecell{0.0431\\(0.0021)} & 0.1053 & 0.1138 \\
& 7 & \makecell{\textbf{0.0769}\\(0.0035)} & \makecell{0.0794\\(0.0060)} & \makecell{0.1121\\(0.0185)} & \makecell{0.0801\\(0.0024)} & 0.1635 & 0.1961 \\
\midrule
\multirow{3}{*}{\textbf{DJI}}
& 1 & \makecell{0.0236\\(0.0040)} & \makecell{\textbf{0.0188}\\(0.0014)} & \makecell{0.0564\\(0.0027)} & \makecell{0.0209\\(0.0032)} & 0.0767 & 0.0925 \\
& 3 & \makecell{0.0392\\(0.0043)} & \makecell{\textbf{0.0380}\\(0.0028)} & \makecell{0.0691\\(0.0029)} & \makecell{0.0445\\(0.0023)} & 0.0979 & 0.1202 \\
& 7 & \makecell{0.0834\\(0.0046)} & \makecell{0.0866\\(0.0055)} & \makecell{0.0988\\(0.0052)} & \makecell{\textbf{0.0782}\\(0.0059)} & 0.1470 & 0.1917 \\
\midrule
\multirow{3}{*}{\textbf{NASDAQ}}
& 1 & \makecell{0.0353\\(0.0071)} & \makecell{0.0324\\(0.0020)} & \makecell{0.0841\\(0.0033)} & \makecell{\textbf{0.0308}\\(0.0017)} & 0.1342 & 0.1446 \\
& 3 & \makecell{\textbf{0.0604}\\(0.0062)} & \makecell{0.0624\\(0.0047)} & \makecell{0.1116\\(0.0033)} & \makecell{0.0652\\(0.0017)} & 0.1659 & 0.1835 \\
& 7 & \makecell{\textbf{0.1179}\\(0.0055)} & \makecell{0.1179\\(0.0079)} & \makecell{0.1527\\(0.0035)} & \makecell{0.1192\\(0.0031)} & 0.2594 & 0.3136 \\
\bottomrule
\end{tabular}
}
\captionsetup{justification=raggedright, singlelinecheck=false}
\caption*{\scriptsize Entries are mean (standard deviation) over 20 seeds where applicable. Deterministic GARCH benchmarks omit standard deviations. `bm\_GARCH\_LSTM' denotes the benchmark pipeline-based hybrid model, where the GARCH component and the LSTM network are trained sequentially rather than jointly.}
\end{table}

The out-of-sample $R^2$ results in Table.\ref{tab:metric_r2} reinforce these observations. This metric evaluates how well the model's predictions match the observed test data on average, providing an interpretable measure of explanatory power in an out-of-sample context.  For all datasets, the GARCH-RNN, especially GARCH-GRU, yields the highest or near-highest $R^2$ values, suggesting that it captures a larger proportion of the variance in realized volatility compared to competing approaches. The advantage is especially pronounced for the S\&P 500, where the $R^2$ remains above 0.87 at horizon 1 and above 0.49 even at horizon 7, demonstrating strong explanatory power and temporal stability. The Transformer models also exhibit solid performance, though the GARCH-RNN’s ability to retain information over longer sequences translates into slightly higher predictability for extended horizons. In contrast, the traditional GARCH and GJR-GARCH models show severely degraded $R^2$ values, underscoring their limited capacity to generalize beyond short-term conditional variance estimation. This gap highlights the necessity of neural augmentations for effectively modeling high-dimensional and regime-dependent volatility dynamics.

% ===================== R^2 (placeholder) =====================
\begin{table}[htbp]
\centering
\caption{Forecasting Performance in Normal-Period (Metric: $R^2$)}
\label{tab:metric_r2}
\setlength{\tabcolsep}{5pt}
\resizebox{\textwidth}{!}{
\begin{tabular}{llcccccc}
\toprule
\textbf{Dataset} & \textbf{Horizon} &
\textbf{GARCH\_GRU} & \textbf{GARCH\_LSTM} &
\textbf{bm\_GARCH\_LSTM} & \textbf{Transformer} &
\textbf{GARCH(1,1)} & \textbf{GJR-GARCH} \\
\midrule
\multirow{3}{*}{\textbf{S\&P 500}} & 1 & \makecell{\textbf{0.8793}\\(0.0057)} & \makecell{0.8532\\(0.0134)} & \makecell{0.6448\\(0.0086)} & \makecell{0.8744\\(0.0078)} & -415.9270 & -517.5718 \\
& 3 & \makecell{\textbf{0.7618}\\(0.0147)} & \makecell{0.7527\\(0.0172)} & \makecell{0.5242\\(0.0127)} & \makecell{0.7148\\(0.0136)} & 0.3049 & 0.2493 \\
& 7 & \makecell{\textbf{0.4900}\\(0.0230)} & \makecell{0.4731\\(0.0398)} & \makecell{0.2557\\(0.1228)} & \makecell{0.4686\\(0.0161)} & -0.0843 & -0.3007 \\
\midrule
\multirow{3}{*}{\textbf{DJI}} & 1 & \makecell{0.8285\\(0.0295)} & \makecell{\textbf{0.8630}\\(0.0104)} & \makecell{0.5894\\(0.0198)} & \makecell{0.8483\\(0.0234)} & -412.3757 & -519.1085 \\
& 3 & \makecell{0.7140\\(0.0312)} & \makecell{\textbf{0.7225}\\(0.0207)} & \makecell{0.4956\\(0.0214)} & \makecell{0.6755\\(0.0164)} & 0.2865 & 0.1243 \\
& 7 & \makecell{0.3877\\(0.0335)} & \makecell{0.3643\\(0.0407)} & \makecell{0.2740\\(0.0384)} & \makecell{\textbf{0.4262}\\(0.0434)} & -0.0784 & -0.4064 \\
\midrule
\multirow{3}{*}{\textbf{NASDAQ}} & 1 & \makecell{0.8347\\(0.0333)} & \makecell{0.8484\\(0.0095)} & \makecell{0.6067\\(0.0155)} & \makecell{\textbf{0.8559}\\(0.0082)} & -433.5734 & -520.2138 \\
& 3 & \makecell{\textbf{0.7176}\\(0.0288)} & \makecell{0.7083\\(0.0221)} & \makecell{0.4780\\(0.0157)} & \makecell{0.6950\\(0.0082)} & 0.2232 & 0.1407 \\
& 7 & \makecell{0.4457\\(0.0258)} & \makecell{\textbf{0.4458}\\(0.0369)} & \makecell{0.2817\\(0.0166)} & \makecell{0.4399\\(0.0144)} & -0.2197 & -0.4744 \\
\bottomrule
\end{tabular}
}
\captionsetup{justification=raggedright, singlelinecheck=false}
\caption*{\scriptsize Entries are mean (standard deviation) over 20 seeds where applicable. Deterministic GARCH benchmarks omit standard deviations. `bm\_GARCH\_LSTM' denotes the pipeline-based hybrid model. Negative OOS $R^2$ values naturally arise under a test-set–mean baseline and simply indicate poorer predictive fit than the mean benchmark.}
\end{table}

The MAE and SMAPE results for the evaluation in 2019 are reported in Appendix.\ref{apd:metric_smape}–\ref{apd:metric_mae} as complementary evidence. These two metrics measure absolute and proportional forecast deviations and therefore, provide an additional perspective alongside the variance-based measures emphasized in the main text. As shown in the appendix tables, the GARCH-RNN models remain competitive under both metrics, although the Transformer occasionally achieves slightly lower errors due to its ability to capture very local patterns in a relatively stable volatility environment. In normal-market conditions, however, variance-based measures such as MSE and out-of-sample $R^2$ are more informative and discriminative, which motivates their use as the primary metrics for the analysis in 2019. By contrast, during the 2020 stress period (Covid-pandemic), characterized by large, abrupt, and highly heterogeneous volatility movements, MAE and SMAPE provide more reliable assessments of absolute and relative forecast accuracy. Accordingly, the subsequent stress-period analysis places greater emphasis on these two metrics.

\begin{figure*}[t]
    \centering
    % Row 1
    \begin{subfigure}[t]{0.49\textwidth}
        \centering
    \includegraphics[width=\linewidth]{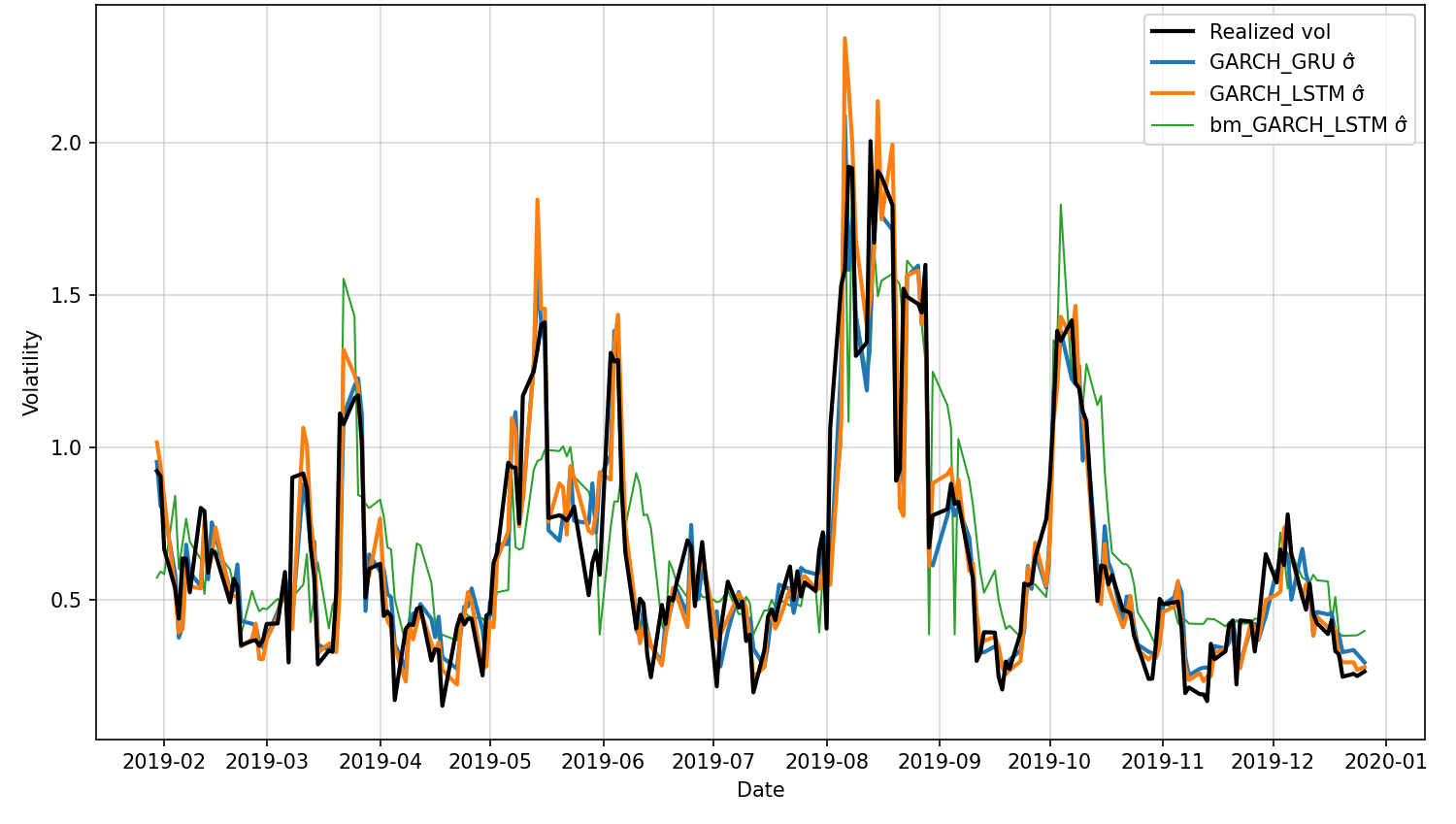}
        \caption{bm\_GARCH\_LSTM vs GARCH\_RNN}
        \label{subfig:bm-lstm}
    \end{subfigure}
    \hfill
    \begin{subfigure}[t]{0.49\textwidth}
        \centering
        \includegraphics[width=\linewidth]{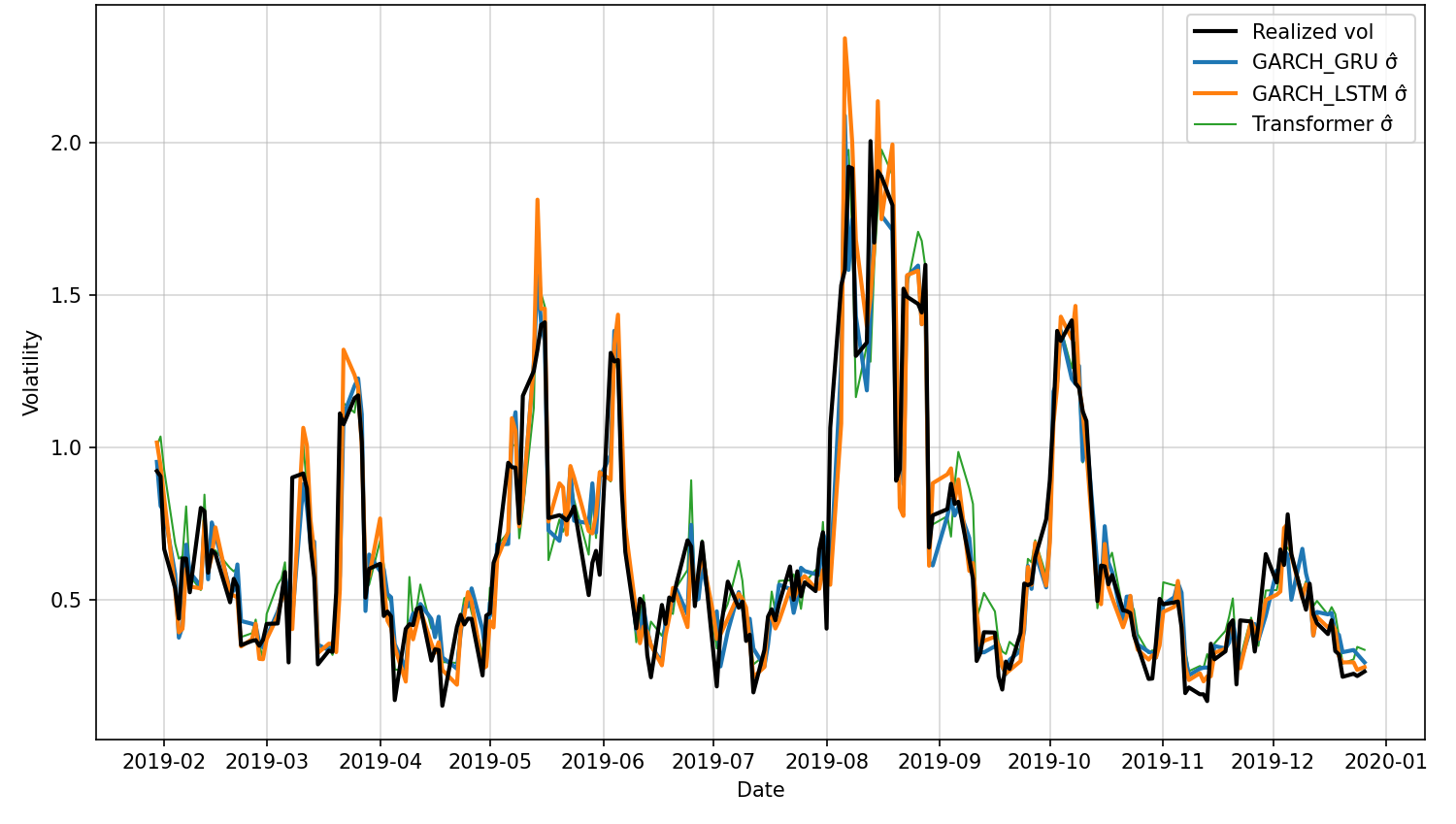}
        \caption{Transformer vs GARCH\_RNN}
        \label{subfig:transformer}
    \end{subfigure}

    \vspace{0.75em}

    % Row 2
    \begin{subfigure}[t]{0.49\textwidth}
        \centering
        \includegraphics[width=\linewidth]{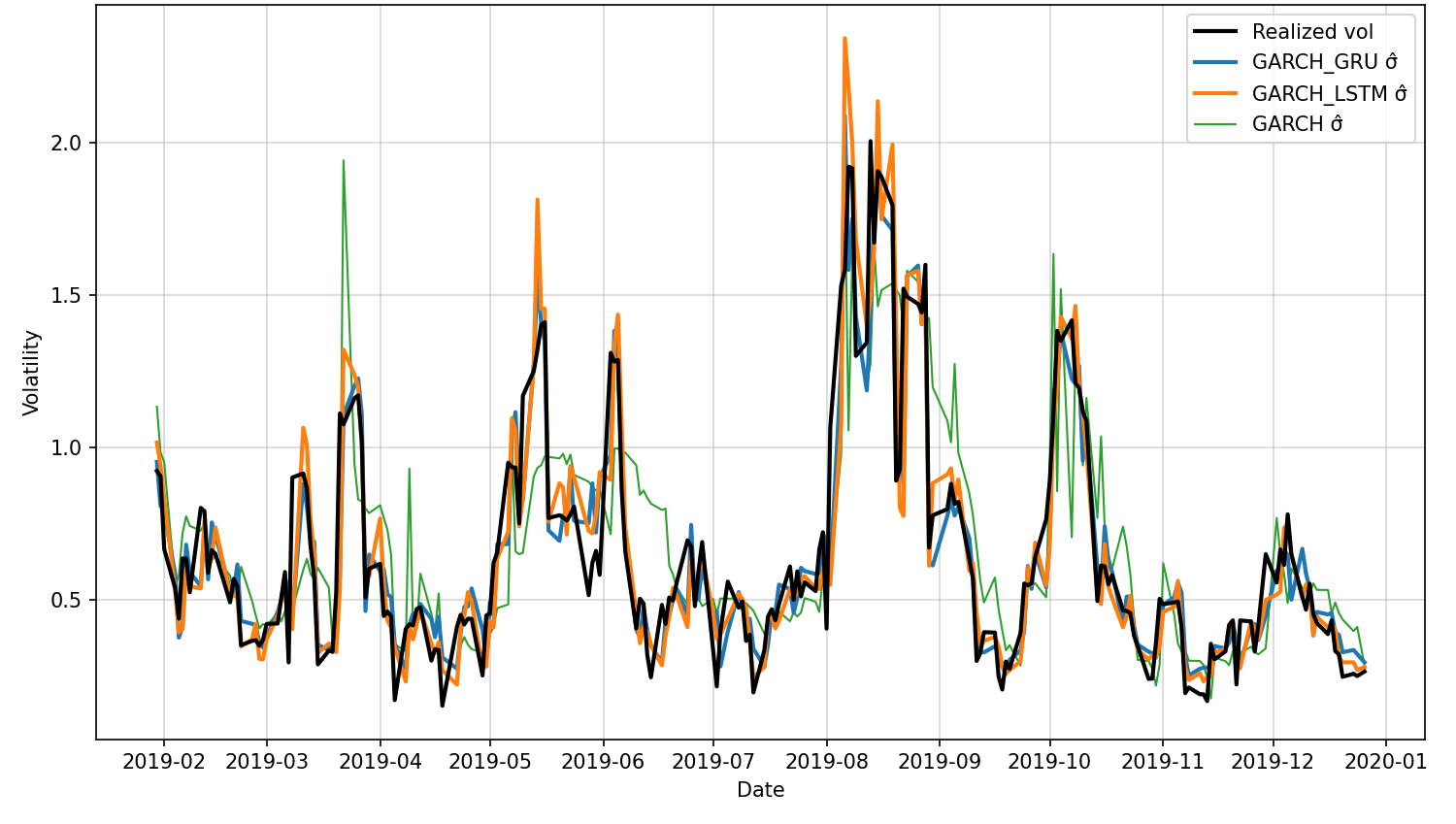}
        \caption{GARCH vs GARCH\_RNN}
        \label{subfig:garch}
    \end{subfigure}
    \hfill
    \begin{subfigure}[t]{0.49\textwidth}
        \centering
        \includegraphics[width=\linewidth]{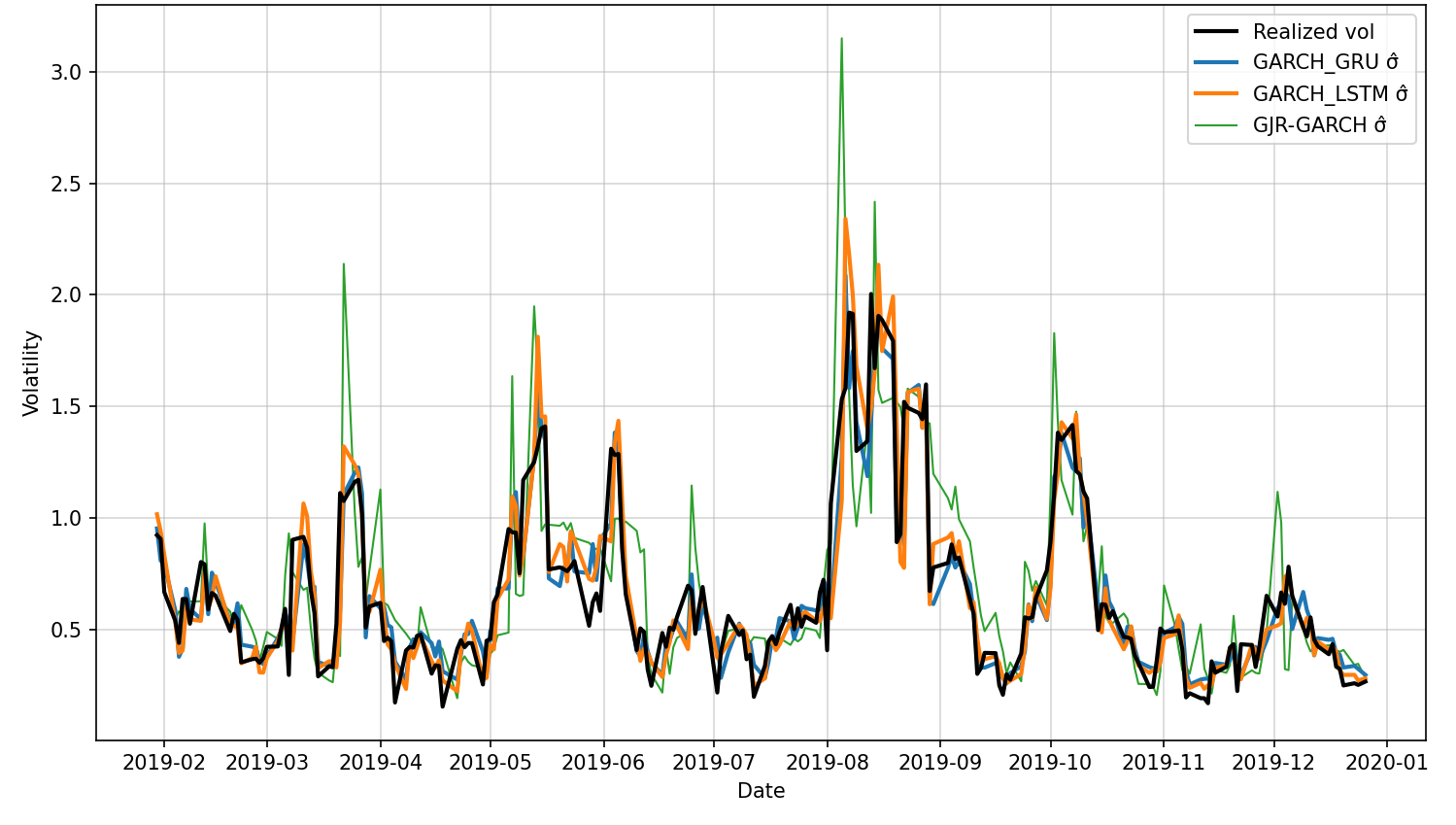}
        \caption{GJR–GARCH vs GARCH\_RNN}
        \label{subfig:gjr}
    \end{subfigure}

    \caption{Response comparison to realized volatility. 
    Each panel overlays realized volatility (black) with the GARCH\_GRU forecast (blue), GARCH\_LSTM forecast (orange) and a comparator model (green).}
    \label{fig:resp-comparison-aug-oct}
\end{figure*}
To complement the quantitative accuracy metrics, we also examine the time-series behavior of the forecasts. Figure.\ref{fig:resp-comparison-aug-oct} plots the predicted volatility paths of GARCH-GRU, GARCH-LSTM, and all benchmark models against realized volatility over the 2019 test window. Both integrated GARCH-RNN models visually track the realized series remarkably well, with close alignment to the timing and magnitude of major volatility swings. This qualitative evidence reinforces the earlier numerical results: the embedded GARCH dynamics allow both recurrent architectures to adapt quickly to changing market conditions.
Across all four panels in Figure.\ref{fig:resp-comparison-aug-oct}, the GARCH-GRU (blue) in particular aligns more promptly with abrupt movements in realized volatility (black). During the early-August volatility surge, GARCH-GRU rises to the new level within one to two days and subsequently relaxes almost immediately as volatility recedes. GARCH-LSTM (orange) exhibits a similarly strong fit, though with slightly smoother transitions that occasionally delay its turning-point response. In contrast, the classical GARCH and GJR-GARCH models (green in panels \ref{subfig:garch}–\ref{subfig:gjr}) display slower mean-reverting adjustments and frequent overshooting, consistent with their rigid parametric structure.
Relative to neural baselines, the difference is equally notable. The two-stage hybrid (bm\_GARCH\_LSTM) shows visibly noisier and less synchronized movements, failing to capture the sharp August and October spikes. The Transformer baseline tracks medium-range fluctuations reasonably well but exhibits noticeable lag at several turning points, such as the late-September trough and early-October jump, where both GARCH-GRU and GARCH-LSTM adjust more quickly and with fewer oscillations. These visual comparisons corroborate that the advantage of the proposed models stems not only from lower error statistics but also from their ability to respond dynamically and coherently to regime shifts in volatility.

%================ High Volatility Subsample ================
\begin{table}[H]
\centering
\begin{threeparttable}
\caption{Volatility Forecasting Performance High-Volatility Subsample of Normal-Period($q=0.9$)}
\label{tab:high_vol_subsample}
\setlength{\tabcolsep}{8pt}
\renewcommand{\arraystretch}{1.2}
\begin{tabular}{lccc}
\toprule
\textbf{Model} & \textbf{MSE} & \textbf{MAE} & \textbf{SMAPE}\\
%($\hat{\sigma}_t, \sigma_t$) \\
\midrule
%GARCH-GRU        & \textbf{1.9595} & \textbf{0.0338} & \textbf{0.5600} \\
%GARCH-LSTM       & 2.0099 & 0.0530 & 0.5359 \\
%bm-GARCH-LSTM    & 2.0791 & 0.0918 & 0.3329 \\
%Transformer      & 1.9711 & 0.0386 & 0.5362 \\
%GARCH            & 2.2582 & 0.1229 & 0.2709 \\
%GJR-GARCH        & 2.0791 & 0.0918 & 0.3329 \\
GARCH\_GRU        & \textbf{0.0679} & \textbf{0.1982} & \textbf{0.1343} \\
GARCH-LSTM       & 0.1056 & 0.2380 & 0.1589 \\
bm\_GARCH-LSTM    & 0.1355 & 0.2958 & 0.2127 \\
Transformer      & 0.0753 & 0.2027 & 0.1398 \\
GARCH            & 0.1896 & 0.3524 & 0.2661 \\
GJR-GARCH        & 0.2472 & 0.3829 & 0.2447 \\
\bottomrule
\end{tabular}
\vspace{-10pt}
\caption*{\scriptsize Results are based on 1-day-ahead forecasts using the S\&P 500 in the 2019 normal-period test sample.}
\end{threeparttable}
\end{table}

Additional perspective is provided by Table.\ref{tab:high_vol_subsample}, which evaluates model performance within a high-volatility subsample (defined by the top 10\% of realized volatility observations). This setting mimics market stress periods, where accurate volatility prediction becomes most relevant for financial stability and risk control. The GARCH-GRU remains the strongest performer across all three metrics (MSE, MAE, and SMAPE), indicating that it not only captures the magnitude of volatility spikes more accurately but also adapts more rapidly to the abrupt, nonlinear fluctuations characteristic of turbulent markets.

For completeness, we also compare the unified GARCH–GRU/LSTM (GARCH-RNN) with standard GRU/LSTM model trained under the same settings. While the GARCH-RNN does not consistently outperform the plain NN across all datasets and horizons, their performance differences are generally small. This indicates that incorporating the GARCH mechanism preserves the NN’s predictive capability while providing stronger theoretical interpretability and stability across different market conditions.

% ===================== GARCH parameters from SP500 =====================
\begin{table}[H]
\centering
\begin{threeparttable}
\caption{Estimated GARCH Parameters for S\&P 500}
\label{tab:garch_params_sp500}

\setlength{\tabcolsep}{6pt}
\renewcommand{\arraystretch}{1.1}
\begin{tabular}{llcccc}
\toprule
\textbf{Model} & \textbf{Horizon} & $\boldsymbol{\omega}$ & $\boldsymbol{\alpha}$ & $\boldsymbol{\beta}$ & $\boldsymbol{\alpha+\beta}$ \\
\midrule
\multirow{3}{*}{GARCH\_GRU} 
 & $h=1$ &    0.0244    &    0.1063    &   0.8637   &  0.9700 \\
 & $h=3$ &    0.0300    &    0.1102    &   0.8598   & 0.9700 \\
 & $h=7$ &    0.0223    &    0.1106    &    0.8592  &  0.9689 \\
\midrule
\multirow{3}{*}{GARCH\_LSTM} 
 & $h=1$ &    0.0260    &    0.0861    &    0.8839   & 0.9700 \\
 & $h=3$ &    0.0214    &    0.0863    &    0.8837   & 0.9700 \\
 & $h=7$ &    0.0205    &    0.0863    &    0.8836   & 0.9699 \\
\midrule
\multirow{3}{*}{GARCH(1,1)} 
 & $h=1$ &    0.0940    &    0.1328    &    0.7125   & 0.8453 \\
 & $h=3$ &    0.0969    &    0.1317    &    0.7135   & 0.8452 \\
 & $h=7$ &    0.1003    &    0.1351    &    0.7188   & 0.8539 \\
\midrule
\multirow{3}{*}{GJR-GARCH} 
 & $h=1$ &    0.1069    &    0.1135    &    0.5660   & 0.6795 \\
 & $h=3$ &    0.1139    &    0.1135    &    0.5611   & 0.6746 \\
 & $h=7$ &    0.1275    &    0.1135    &    0.5550   & 0.6685 \\
\bottomrule
\end{tabular}
\begin{minipage}{\linewidth}
\scriptsize
For GJR-GARCH, the leverage parameter $\gamma$ is estimated but omitted here for comparability. 
\end{minipage}
\end{threeparttable}
\end{table}

A further advantage of the proposed GARCH-RNN architecture lies in its ability to preserve interpretable GARCH dynamics while jointly learning nonlinear recurrent structure, as reflected in Eq.\ref{interpretability}. Table.\ref{tab:garch_params_sp500} reports the estimated $\omega$, $\alpha$, and $\beta$ coefficients extracted from the embedded GARCH layer for the S\&P 500. 
Both GARCH-GRU and GARCH-LSTM produce economically plausible coefficients. The shock response $\alpha$ remains moderate ( $\approx 0.09-0.11$ ), the persistence parameter $\beta$ remains high $(\approx 0.86-0.88)$, and the resulting $\alpha +\beta \approx 0.97 $ closely matches well-documented persistence in equity volatility. Notably, these parameter values remain stable across forecasting horizons, indicating that unified training does not distort the underlying GARCH dynamics but instead complements them with nonlinear recurrent structure.
Compared with classical GARCH and GJR-GARCH estimates, the embedded GARCH layers yield lower unconditional variance ($\omega$) and substantially higher persistence, consistent with the smoother and more adaptive volatility transitions observed in the time-series plots. These results demonstrate that the GARCH-RNN models retain interpretable GARCH behavior while benefiting from neural gating mechanisms, helping explain their superior predictive accuracy across the evaluated horizons and datasets.

\subsubsection{Stress-Period Evaluation (2020 COVID Pandemic)}
Having established model performance under normal market conditions, we now examine the behavior of all models during the 2020 COVID-19 period, a regime characterized by extreme volatility, rapid structural breaks, and unusually persistent shocks. This period provides a stringent test of robustness, as forecasting errors typically amplify when volatility levels rise sharply and shift unpredictably. In this setting, absolute- and scale-adjusted deviation measures such as MAE and SMAPE become particularly informative, since they directly quantify the magnitude and proportional severity of prediction errors during large fluctuations.

%===================== MAE =====================
\begin{table}[H]
\centering
\caption{Forecasting Performance in Stress-Period (Metric: MAE)}
\label{tab:metric_mae_extended}
\setlength{\tabcolsep}{5pt}
\resizebox{\textwidth}{!}{
\begin{tabular}{llcccccc}
\toprule
\textbf{Dataset} & \textbf{Horizon} &
\textbf{GARCH\_GRU} & \textbf{GARCH\_LSTM} &
\textbf{bm\_GARCH\_LSTM} & \textbf{Transformer} &
\textbf{GARCH(1,1)} & \textbf{GJR-GARCH} \\
\midrule
\multirow{3}{*}{\textbf{S\&P 500}}
& 1 & \makecell{0.3408\\(0.0127)} & \makecell{\textbf{0.3360}\\(0.0141)} & \makecell{0.4255\\(0.0131)} & \makecell{0.4358\\(0.0350)} & 0.4441 & 0.4764 \\
& 3 & \makecell{\textbf{0.4451}\\(0.0117)} & \makecell{0.4632\\(0.0185)} & \makecell{0.5048\\(0.0319)} & \makecell{0.5293\\(0.0206)} & 0.5290 & 0.5552 \\
& 7 & \makecell{0.6361\\(0.0147)} & \makecell{\textbf{0.6358}\\(0.0244)} & \makecell{0.7011\\(0.1014)} & \makecell{0.6622\\(0.0121)} & 0.6691 & 0.7151 \\
\midrule
\multirow{3}{*}{\textbf{DJI}}
& 1 & \makecell{0.4581\\(0.0269)} & \makecell{\textbf{0.3932}\\(0.0195)} & \makecell{0.4981\\(0.0199)} & \makecell{0.5255\\(0.0267)} & 0.4570 & 0.4842 \\
& 3 & \makecell{0.5295\\(0.0215)} & \makecell{\textbf{0.5121}\\(0.0319)} & \makecell{0.5613\\(0.0177)} & \makecell{0.6268\\(0.0222)} & 0.5507 & 0.5682 \\
& 7 & \makecell{0.7054\\(0.0157)} & \makecell{0.7507\\(0.0203)} & \makecell{0.7298\\(0.0163)} & \makecell{0.7446\\(0.0108)} & \textbf{0.6945} & 0.7443 \\
\midrule
\multirow{3}{*}{\textbf{NASDAQ}}
& 1 & \makecell{0.3939\\(0.0225)} & \makecell{\textbf{0.3682}\\(0.0147)} & \makecell{0.4833\\(0.0074)} & \makecell{0.4375\\(0.0347)} & 0.4857 & 0.4911 \\
& 3 & \makecell{0.5013\\(0.0176)} & \makecell{\textbf{0.4781}\\(0.0167)} & \makecell{0.5313\\(0.0195)} & \makecell{0.5424\\(0.0164)} & 0.5827 & 0.5791 \\
& 7 & \makecell{0.6955\\(0.0094)} & \makecell{\textbf{0.6703}\\(0.0120)} & \makecell{0.7029\\(0.0200)} & \makecell{0.6868\\(0.0086)} & 0.7387 & 0.7568 \\
\bottomrule
\end{tabular}
}
\captionsetup{justification=raggedright, singlelinecheck=false}
\caption*{\scriptsize Entries are Stress-period (2020) MAE mean (standard deviation) across 20 seeds where applicable. Deterministic GARCH benchmarks omit standard deviations.}
\end{table}

%===================== SMAPE =====================
\begin{table}[H]
\centering
\caption{Forecasting Performance in Stress-Period (Metric: SMAPE)}
\label{tab:metric_smape_extended}
\setlength{\tabcolsep}{5pt}
\resizebox{\textwidth}{!}{
\begin{tabular}{llcccccc}
\toprule
\textbf{Dataset} & \textbf{Horizon} &
\textbf{GARCH\_GRU} & \textbf{GARCH\_LSTM} &
\textbf{bm\_GARCH\_LSTM} & \textbf{Transformer} &
\textbf{GARCH(1,1)} & \textbf{GJR-GARCH} \\
\midrule
\multirow{3}{*}{\textbf{S\&P 500}}
& 1 & \makecell{\textbf{0.1804}\\(0.0056)} & \makecell{0.1807\\(0.0068)} & \makecell{0.2900\\(0.0063)} & \makecell{0.2221\\(0.0121)} & 0.7180 & 0.7268 \\
& 3 & \makecell{\textbf{0.2482}\\(0.0058)} & \makecell{0.2515\\(0.0057)} & \makecell{0.3229\\(0.0110)} & \makecell{0.2872\\(0.0083)} & 0.3401 & 0.3430 \\
& 7 & \makecell{0.3649\\(0.0065)} & \makecell{\textbf{0.3606}\\(0.0105)} & \makecell{0.4186\\(0.0648)} & \makecell{0.3732\\(0.0059)} & 0.3993 & 0.4156 \\
\midrule
\multirow{3}{*}{\textbf{DJI}}
& 1 & \makecell{0.2070\\(0.0129)} & \makecell{\textbf{0.1850}\\(0.0083)} & \makecell{0.2867\\(0.0065)} & \makecell{0.2415\\(0.0103)} & 0.7056 & 0.7080 \\
& 3 & \makecell{0.2573\\(0.0103)} & \makecell{\textbf{0.2553}\\(0.0133)} & \makecell{0.3153\\(0.0082)} & \makecell{0.3059\\(0.0103)} & 0.3171 & 0.3307 \\
& 7 & \makecell{\textbf{0.3575}\\(0.0082)} & \makecell{0.3772\\(0.0108)} & \makecell{0.3954\\(0.0073)} & \makecell{0.3756\\(0.0062)} & 0.3766 & 0.4028 \\
\midrule
\multirow{3}{*}{\textbf{NASDAQ}}
& 1 & \makecell{0.2035\\(0.0119)} & \makecell{\textbf{0.1907}\\(0.0067)} & \makecell{0.2938\\(0.0054)} & \makecell{0.2141\\(0.0139)} & 0.6737 & 0.6838 \\
& 3 & \makecell{0.2684\\(0.0081)} & \makecell{\textbf{0.2677}\\(0.0093)} & \makecell{0.3326\\(0.0082)} & \makecell{0.2879\\(0.0067)} & 0.3546 & 0.3435 \\
& 7 & \makecell{0.3849\\(0.0058)} & \makecell{\textbf{0.3726}\\(0.0062)} & \makecell{0.4076\\(0.0105)} & \makecell{0.3762\\(0.0056)} & 0.4178 & 0.4170 \\
\bottomrule
\end{tabular}
}
\captionsetup{justification=raggedright, singlelinecheck=false}
\caption*{\scriptsize Entries are Stress-period (2020) SMAPE mean (standard deviation) across 20 seeds where applicable. Deterministic GARCH benchmarks omit standard deviations.}
\end{table}

The results in Tables.\ref{tab:metric_mae_extended}–\ref{tab:metric_smape_extended} demonstrate that the proposed GARCH-RNN (both GARCH-GRU and GARCH-LSTM) framework maintains a clear and substantial advantage over all benchmark models during the COVID-19 stress period. Both MAE and SMAPE increase sharply across all models due to unprecedented market turbulence, yet the GARCH-RNN architectures exhibit markedly smaller error escalations than both classical GARCH specifications and non-embedded neural networks. This confirms that integrating GARCH dynamics directly into the recurrent structure provides essential stability and adaptability under extreme volatility, where conventional econometric models and purely data-driven deep learning methods struggle to accommodate abrupt regime shifts.
Within the GARCH-RNN family, a distinct pattern emerges. GARCH-LSTM achieves the lowest MAE across most datasets and horizons, particularly at the one-day and three-day horizons where shocks propagate rapidly. Its more elaborate gating mechanism appears to provide additional memory depth that becomes advantageous when volatility levels remain elevated for extended periods. However, GARCH-GRU remains highly competitive, and in several settings, especially for the S\&P 500, it attains the best SMAPE values, indicating stronger proportional accuracy when volatility spikes. The performance gap between the two GARCH-RNN variants is small, especially relative to the gap separating both models from the benchmarks, reinforcing the conclusion that the primary performance gains arise from the embedded GARCH–neural interaction itself rather than from the choice of recurrent cell.

Benchmark models deteriorate much more severely under stress. The two-stage hybrid (bm\_GARCH\_LSTM) performs considerably worse than either of the embedded GARCH-RNN models, underscoring the importance of joint end-to-end learning for capturing rapidly shifting volatility dynamics. Transformer models, despite their strong performance in benign environments, experience notable degradation, revealing limitations of purely data-driven architectures when confronted with large structural breaks. Classical GARCH and GJR-GARCH models exhibit the weakest performance overall, reflecting their inability to adapt to sudden persistence changes and nonlinear amplification common in crisis periods.

%Taken together, the evidence suggests a regime-dependent pattern within the GARCH-NN family: GARCH-GRU is the superior model in stable or moderately volatile environments, due to its responsiveness and efficient gating, whereas GARCH-LSTM gains a relative advantage under extreme stress, where its additional memory depth better captures prolonged turbulence. Importantly, both models significantly outperform traditional econometric specifications and non-embedded neural architectures, validating the robustness and flexibility of the proposed GARCH-NN framework in adverse market conditions.

\begin{table}[H]
\centering
\caption{Average Training Time per Epoch Comparison}
\label{tab:runtime_avg}
\setlength{\tabcolsep}{8pt}
\renewcommand{\arraystretch}{1.25}
\begin{threeparttable}
\begin{tabular}{lc}
\toprule
\textbf{Model} & \textbf{Avg. Time per Epoch (s)}\\
\midrule
GARCH\_GRU  & 665.78 \\
GARCH\_LSTM & 1838.81 \\
\bottomrule
\end{tabular}
%\begin{tablenotes}
\scriptsize
The reported average training times are aggregated across all forecasting horizons, datasets, and random seeds (20 runs in total).
%\end{tablenotes}
\end{threeparttable}
\end{table}

Furthermore, to evaluate the computational efficiency, we measured the average training time per epoch for both GARCH-GRU and GARCH-LSTM models across 20 runs with different random seeds as shown in Table.\ref{tab:runtime_avg}. Aggregating the results across all forecasting horizons, datasets, and seeds, the GARCH-GRU model exhibits an average epoch time of 665.78 seconds, while the GARCH-LSTM requires 1838.81 seconds. This implies that, on average, the GARCH-GRU architecture trains approximately 2.76 times faster than its LSTM counterpart. The superior efficiency of GARCH-GRU arises from its simpler gating structure, which reduces the number of recurrent parameters and matrix multiplications at each time step. Consequently, the GRU-based hybrid model delivers a substantial reduction in computational cost without sacrificing predictive accuracy, making it a more scalable and time-efficient framework for multi-horizon or high-frequency volatility forecasting applications.

Overall, these results consistently demonstrate that coupling a recurrent architecture with a GARCH component enables the model to leverage both the parametric interpretability of volatility dynamics and the nonlinear representational power of deep networks. Among all tested frameworks, the GARCH-GRU offers superior performance in stable or moderately volatile markets, while GARCH-LSTM gains a relative edge under extreme stress, yet both significantly outperform all benchmarks. These findings validate the GARCH-RNN framework as a flexible, interpretable, and empirically powerful approach to volatility forecasting across diverse market regimes.

%=============================================================
\section{Applications}
One of the most important practical applications of volatility forecasting lies in financial risk management, particularly in the computation of Value-at-Risk (VaR). VaR is a widely adopted quantitative risk measure that estimates the maximum potential loss of a portfolio or asset over a specified time horizon at a given confidence level. It is extensively used by financial institutions, regulators, and portfolio managers to assess market risk exposure, allocate capital, and comply with regulatory requirements such as those imposed under the Basel Accords.

Formally, for a return series $\{r_t\}$, the one-step-ahead VaR at significance level $\alpha \in (0,1)$ is defined as the conditional quantile of the return distribution such that the probability of a loss exceeding this level is no greater than $\alpha$. We express $\operatorname{VaR}$ on the loss scale, so that $\operatorname{VaR}_{t+1}^{(\alpha)}>0$ represents the $(1-\alpha)$-quantile of the conditional loss distribution. Under this convention, a VaR exceedance occurs whenever the realized return falls below the negative of this threshold:
\begin{equation}
\mathbb{P}\bigl(r_{t+1} < -\text{VaR}_{t+1}^{(\alpha)} \mid \mathcal{F}_t\bigr) = \alpha,
\end{equation}
where $\mathcal{F}_t$ denotes the information set at time $t$.
Assuming a constant conditional mean $\mu$, we compute VaR by combining the model’s volatility forecast with a parametric quantile based on the standardized return distribution. In this study, we adopt a Student-\emph{t} specification for standardized residuals, which better captures the heavy tails commonly observed in equity returns. Let the standardized residuals be
\[
\hat{\varepsilon}_t = \frac{r_t - \mu}{\sigma_t},
\]
and let $q_\alpha^{(t)}$ denote the $\alpha$-quantile of the fitted Student-\emph{t} distribution. The one-step-ahead VaR is then given by
\begin{equation}
\text{VaR}_{t+1}^{(\alpha)} = -\mu - q_\alpha^{(t)} \,\hat{\sigma}_{t+1},
\end{equation}
where $\hat{\sigma}_{t+1}$ is the volatility forecast obtained from the model. 

To evaluate the quality of the VaR forecasts, we first employ coverage-based diagnostics: the violation count and the violation ratio. A violation occurs when the actual return falls below the estimated VaR threshold:
\begin{equation}
\mathbb{I}\left\{r_{t+1}<-\operatorname{VaR}_{t+1}^{(\alpha)}\right\}=1
\end{equation}
indicating an underestimation of the true downside risk. The violation count is defined as:
\begin{equation}
\text { Violation Ratio }=\frac{\text { Number of Violations }}{\text { Total Number of Forecasts }}
\end{equation}
and is compared to the nominal level $\alpha$ to assess whether the model underestimates or overestimates risk.

%Pinball loss
In addition to coverage-based evaluation, we further assess the accuracy of the predicted quantiles using the Pinball loss (also known as quantile loss). Unlike the violation ratio, which measures calibration at a specific confidence level, the Pinball loss provides a strictly consistent scoring rule that evaluates both the sharpness and reliability of quantile forecasts. For a realized return $r_{t+1}$ and the predicted quantile $\widehat{\operatorname{VaR}}_{t+1}^{(\alpha)}$, the Pinball loss is defined as

\begin{equation}
\mathcal{L}_\alpha\left(r_{t+1}, \widehat{\operatorname{VaR}}_{t+1}^{(\alpha)}\right)= \begin{cases}\alpha \cdot\left(r_{t+1}+\widehat{\operatorname{VaR}}_{t+1}^{(\alpha)}\right), & \text { if } r_{t+1} \geq-\widehat{\operatorname{VaR}}_{t+1}^{(\alpha)} \\ (1-\alpha) \cdot\left(-r_{t+1}-\widehat{\operatorname{VaR}}_{t+1}^{(\alpha)}\right), & \text { otherwise. }\end{cases}
\end{equation}
Lower values of $\mathcal{L}_\alpha$ indicate more accurate and better-calibrated quantile forecasts, making the Pinball loss a natural complement to traditional violation-based backtesting in our VaR evaluation.

\begin{figure}[!htbp]
  \centering
\includegraphics[width=0.8\columnwidth]{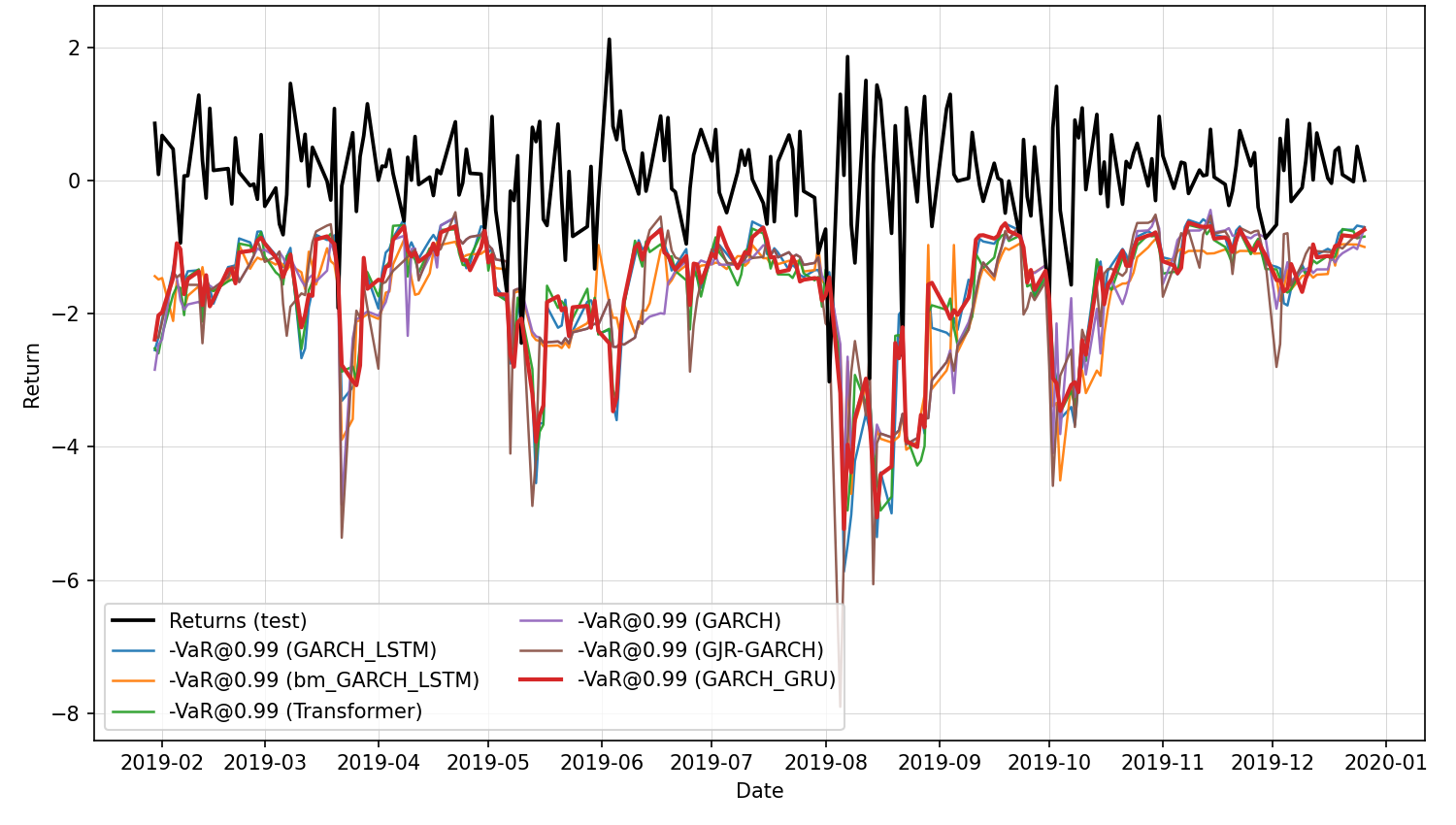}
  \caption{Actual Returns \& VaR Forecasts at 99\% Confidence Level on S\&P 500}
  \label{fig:VaR}
\end{figure}

\begin{table}[H]
\centering
\setlength{\tabcolsep}{6pt}
\renewcommand{\arraystretch}{1.25}
%\begin{threeparttable}
\caption{Volatility and Tail-Risk Evaluation}
\label{tab:oos_risk}
%================ Panel B ================
\begin{threeparttable}
\begin{tabular}{lccc}
\toprule
\multicolumn{4}{c}{\textbf{VaR Backtesting ($\alpha=0.01$)}}\\
\midrule
Model & Violation Rate & Violations & Pinball\_loss \\
\midrule
GARCH\_GRU      & \textbf{0.0261} & \textbf{6} & 0.0296 \\
GARCH\_LSTM     & 0.0348 & 8 & 0.0326 \\
bm\_GARCH\_LSTM & 0.0261 & 6 & 0.0321 \\
Transformer     & 0.0304 & 7 & \textbf{0.0286} \\
GARCH           & 0.0261 & 6 & 0.0361 \\
GJR-GARCH       & 0.0261 & 6 & 0.0330 \\
\bottomrule
\end{tabular}

\scriptsize
The VaR backtest is based on 230 one-day-ahead forecasts over the 2019 test window.
\end{threeparttable}
\end{table}

\begin{comment}
\vspace{1em}
%================ Panel D ================
\begin{tabular}{lcc}
\toprule
\multicolumn{3}{c}{\textbf{Panel D: Diebold–Mariano Tests (QLIKE Loss)}}\\
\midrule
Model A vs B & DM stat & p-value\\
\midrule
GARCH\_GRU vs GARCH\_LSTM       & -1.0303 & 0.3029\\
GARCH\_GRU vs bm\_GARCH\_LSTM   & -4.4205 & \textbf{9.85e-06}\\
GARCH\_GRU vs Transformer        &  1.9664 & 0.04925\\
GARCH\_GRU vs GARCH              & -4.6527 & \textbf{3.28e-06}\\
GARCH\_GRU vs GJR-GARCH          & -4.4205 & \textbf{9.85e-06}\\
\bottomrule
\end{tabular}
\end{comment}

Figure.\ref{fig:VaR} visualizes the one-day-ahead 99\% VaR forecasts from all models against the realized S\&P 500 returns in 2019. A correctly calibrated model should generate a VaR envelope that consistently lies below (i.e., more negative than) the realized return series except on rare occasions corresponding to true tail events. Both GARCH-RNN models, particularly GARCH-GRU, produce VaR paths that closely and coherently track the depth and timing of downside risk. Their VaR curves widen promptly around the major loss episodes in August and October 2019, reflecting a rapid increase in predictive tail thickness as volatility rises. By contrast, the classical GARCH-type models show noticeably slower adjustments and tend to underestimate risk immediately before sharp drawdowns, resulting in several clustered exceedances. The pipeline hybrid (bm\_GARCH\_LSTM) exhibits noisier and less stable VaR paths, again indicating weaker joint modeling of volatility and tail behavior. The Transformer captures some local loss patterns but displays occasional under-reactions around abrupt volatility spikes, consistent with its limitations under nonstationary dynamics. Overall, the visual evidence confirms that embedding GARCH structure within the recurrent cell enables both GARCH-GRU and GARCH-LSTM to generate more coherent and timely tail-risk estimates.

Table.\ref{tab:oos_risk} further quantifies tail-risk performance using standard VaR backtesting metrics at the 99\% level. A well-calibrated model should exhibit a violation rate close to the nominal 1\% level, a small number of exceedances, and low Pinball loss values that reflect sharp and well-aligned quantile forecasts. Among all models, GARCH-GRU delivers the most balanced and stable performance, with six violations and a violation rate of 0.0261, while maintaining tight Pinball loss scores. Its consistent behavior across both metrics indicates strong conditional tail calibration and reliable responsiveness to latent volatility dynamics. The GARCH-LSTM variant also performs competitively but shows slightly higher violation frequency and weaker quantile sharpness, suggesting that its additional gating depth does not always translate into superior tail alignment during moderate volatility regimes.
The two-stage hybrid (bm\_GARCH\_LSTM) matches the violation rate of GARCH-GRU but suffers from noticeably higher Pinball loss, reinforcing the advantage of unified learning over sequential pipeline designs. The Transformer achieves the lowest Pinball loss among all models, indicating highly sharp quantile prediction. However, its higher violation count reveals underestimation of tail intensity, which is an imbalance commonly seen in purely data-driven architectures lacking explicit volatility structure. Classical GARCH and GJR-GARCH models attain acceptable violation frequencies but exhibit substantially higher Pinball losses, confirming limited flexibility by their parametric structure and reduced capacity to match the sharpness and adaptability achieved by the neural and hybrid architectures in quantile forecasting.

In short, these findings demonstrate that the GARCH-RNN models, especially GARCH-GRU, deliver superior tail-risk performance. Their VaR forecasts respond rapidly and consistently to changes in market volatility, producing quantiles that track downside risk with appropriate depth and timing. The backtesting results further confirm robust calibration and quantile sharpness. These properties make the proposed GARCH-RNN framework particularly suitable for practical risk-management tasks, including VaR/ES estimation, stress testing, and capital-adequacy assessment. Overall, the results thus affirm the model’s capacity to bridge the gap between statistical volatility modeling and actionable financial risk control.

\section{Conclusion and Future Work}

This study contributes to the advancement of financial volatility modeling by proposing a novel hybrid architecture that unifies traditional econometric structure with modern deep learning techniques. Specifically, we introduce two integrated variants, GARCH-GRU and GARCH-LSTM, in which GARCH(1,1) volatility dynamics are embedded directly into the recurrent computation via a dedicated multiplicative gate. This construction enables the models to capture volatility clustering, long memory, and nonlinear temporal dependence within a single unified trainable system.
An important advantage of this design is that it preserves interpretable GARCH parameters ($\omega$, $\alpha$, $\beta$), allowing the models to retain clear economic meaning while benefiting from the representational flexibility of neural networks.

Comprehensive empirical evaluation across major U.S. equity indices (S\&P 500, Dow Jones Industrial Average, NASDAQ) demonstrates that both GARCH–GRU and GARCH–LSTM consistently outperform a wide set of benchmarks, including classical GARCH-family models, pipeline-style hybrids, and neural baselines such as the Transformer. Across multiple risk and accuracy metrics (MSE, $R^2$, MAE, SMAPE), the GARCH-RNN models achieve substantially higher predictive accuracy and exhibit greater stability across forecasting horizons.
Within the proposed family, the GARCH–GRU emerges as the primary and most versatile variant, offering the strongest performance in stable or moderately volatile environments and achieving the best overall accuracy–efficiency tradeoff. It requires less than two-thirds of the training time of the GARCH–LSTM due to its more compact gating structure. The GARCH–LSTM, while more computationally intensive, provides additional memory depth that becomes advantageous in highly turbulent periods, allowing it to match or surpass GARCH–GRU in certain stress-period settings. Nevertheless, the performance gap between these two variants is small compared with the considerable gains they jointly achieve over all benchmarks, confirming that the primary improvements arise from the unified embedding of GARCH dynamics within the recurrent architecture.
Moreover, both models yield stable and economically plausible GARCH parameter estimates, confirming that the unified design retains interpretability and diagnostic transparency which is an advantage rarely available in purely data-driven deep learning models.

The risk-based evaluation further confirms the robustness of the proposed framework. In 99\% VaR backtesting, the GARCH-RNN models deliver lower violation ratios and sharper quantile predictions than classical GARCH specifications. The GARCH-GRU exhibits particularly strong calibration, achieving violation frequencies close to the nominal level while maintaining competitive Pinball losses. These results underscore the framework’s ability to adapt to tail-risk behavior and its suitability for practical applications in risk management, stress testing, and capital adequacy assessment.

Overall, the GARCH-RNN framework demonstrates that a unified integration of econometric volatility structure and neural recurrent dynamics can yield models that are simultaneously interpretable, flexible, and empirically powerful. Unlike traditional hybrid approaches that combine econometric and machine-learning components in a sequential or loosely coupled manner, the proposed unified architecture allows both components to reinforce each other within a coherent modeling system. This combination of interpretability, stability, and nonlinear representational capacity positions the GARCH-RNN family as a bridge between classical financial modeling and modern machine learning.

A remaining limitation of the framework lies in the computational overhead associated with custom recurrent-cell implementations. Although the proposed models have been carefully validated and the demonstrate superior efficiency relative to benchmarks, custom operators remain slower than native deep-learning primitives in large-scale or latency-sensitive applications. Future work will focus on code-level optimization, operator fusion, and hardware-aware implementation strategies to improve runtime performance without altering the unified model structure.

Looking ahead, several promising avenues extend naturally from this work. Incorporating asymmetric GARCH components (EGARCH or GJR-GARCH) into the unified architecture may improve the modeling of leverage effects. Integrating attention mechanisms could further enhance pattern recognition, particularly in noisy or high-frequency environments. Regime-switching or state-space extensions would allow the model to adapt to structural breaks and evolving volatility regimes. These directions hold the potential to enrich the unified econometric–neural design, yielding models that are not only interpretable and scalable but also robust to the increasingly complex dynamics of modern financial markets.\\

\noindent\textbf{Data availability statement:}\

\noindent The data that support the findings of this study are openly available at Yahoo Finance.\\

\noindent\textbf{Declaration of interests statement:}\

\noindent The authors declare that they have no known competing financial interests or personal relationships that could have appeared to influence the work reported in this paper.

\newpage % Ensure the references start on a new page
\footnotesize

\bibliographystyle{chicago}
\bibliography{reference}

\newpage
\appendix
\section*{Appendix - Further Experimental Results}
% ===================== SMAPE (placeholder) =====================
\begin{table}[htbp]
\centering
\caption{Forecasting Performance in Normal-Period (Metric: SMAPE)}
\label{apd:metric_smape}
\setlength{\tabcolsep}{5pt}
\resizebox{\textwidth}{!}{
\begin{tabular}{llcccccc}
\toprule
\textbf{Dataset} & \textbf{Horizon} &
\textbf{GARCH\_GRU} & \textbf{GARCH\_LSTM} &
\textbf{bm\_GARCH\_LSTM} & \textbf{Transformer} &
\textbf{GARCH(1,1)} & \textbf{GJR-GARCH} \\
\midrule
\multirow{3}{*}{\textbf{S\&P 500}} & 1 & \makecell{0.1653\\(0.0060)} & \makecell{0.1603\\(0.0061)} & \makecell{0.2771\\(0.0066)} & \makecell{\textbf{0.1529}\\(0.0064)} & 0.5394 & 0.5623 \\
& 3 & \makecell{\textbf{0.2206}\\(0.0105)} & \makecell{0.2236\\(0.0095)} & \makecell{0.3194\\(0.0080)} & \makecell{0.2336\\(0.0056)} & 0.3868 & 0.3738 \\
& 7 & \makecell{\textbf{0.3056}\\(0.0090)} & \makecell{0.3149\\(0.0104)} & \makecell{0.3815\\(0.0203)} & \makecell{0.3058\\(0.0074)} & 0.4559 & 0.4642 \\
\midrule
\multirow{3}{*}{\textbf{DJI}} & 1 & \makecell{0.1803\\(0.0150)} & \makecell{\textbf{0.1617}\\(0.0078)} & \makecell{0.2806\\(0.0120)} & \makecell{0.1688\\(0.0142)} & 0.5097 & 0.5405 \\
& 3 & \makecell{0.2311\\(0.0124)} & \makecell{\textbf{0.2235}\\(0.0066)} & \makecell{0.3075\\(0.0007)} & \makecell{0.2481\\(0.0078)} & 0.3621 & 0.3754 \\
& 7 & \makecell{0.3169\\(0.0056)} & \makecell{0.3202\\(0.0073)} & \makecell{0.3603\\(0.0060)} & \makecell{\textbf{0.3088}\\(0.0121)} & 0.4203 & 0.4489 \\
\midrule
\multirow{3}{*}{\textbf{NASDAQ}} & 1 & \makecell{0.1734\\(0.0148)} & \makecell{0.1591\\(0.0041)} & \makecell{0.2810\\(0.0106)} & \makecell{\textbf{0.1499}\\(0.0058)} & 0.5400 & 0.5608 \\
& 3 & \makecell{\textbf{0.2249}\\(0.0127)} & \makecell{0.2269\\(0.0074)} & \makecell{0.3253\\(0.0060)} & \makecell{0.2270\\(0.0036)} & 0.3875 & 0.3799 \\
& 7 & \makecell{0.3122\\(0.0066)} & \makecell{0.3148\\(0.0092)} & \makecell{0.3735\\(0.0055)} & \makecell{\textbf{0.3102}\\(0.0044)} & 0.4642 & 0.4583 \\
\bottomrule
\end{tabular}
}
\captionsetup{justification=raggedright, singlelinecheck=false}
\caption*{\scriptsize Results are based on the 2019 out-of-sample evaluation. Reported values denote the mean and standard deviation across 20 seeds where applicable.}
\end{table}

% ===================== MAE =====================
\begin{table}[htbp]
\centering
\caption{Forecasting Performance in Normal-Period (Metric: MAE)}
\label{apd:metric_mae}
\setlength{\tabcolsep}{5pt}
\resizebox{\textwidth}{!}{
\begin{tabular}{llcccccc}
\toprule
\textbf{Dataset} & \textbf{Horizon} &
\textbf{GARCH\_GRU} & \textbf{GARCH\_LSTM} &
\textbf{bm\_GARCH\_LSTM} & \textbf{Transformer} &
\textbf{GARCH(1,1)} & \textbf{GJR-GARCH} \\
\midrule
\multirow{3}{*}{\textbf{S\&P 500}}
& 1 & \makecell{0.0945\\(0.0035)} & \makecell{0.0963\\(0.0044)} & \makecell{0.1757\\(0.0040)} & \makecell{\textbf{0.0890}\\(0.0038)} & 0.2174 & 0.2124 \\
& 3 & \makecell{0.1343\\(0.0059)} & \makecell{0.1389\\(0.0068)} & \makecell{0.2042\\(0.0051)} & \makecell{\textbf{0.1146}\\(0.0037)} & 0.2529 & 0.2519 \\
& 7 & \makecell{\textbf{0.1989}\\(0.0062)} & \makecell{0.2071\\(0.0084)} & \makecell{0.2558\\(0.0176)} & \makecell{0.2000\\(0.0042)} & 0.3125 & 0.3322 \\
\midrule
\multirow{3}{*}{\textbf{DJI}}
& 1 & \makecell{0.1073\\(0.0092)} & \makecell{\textbf{0.0954}\\(0.0047)} & \makecell{0.1780\\(0.0068)} & \makecell{0.0991\\(0.0095)} & 0.2101 & 0.2250 \\
& 3 & \makecell{0.1423\\(0.0078)} & \makecell{\textbf{0.1390}\\(0.0052)} & \makecell{0.1970\\(0.0042)} & \makecell{0.1536\\(0.0050)} & 0.2396 & 0.2601 \\
& 7 & \makecell{0.2098\\(0.0050)} & \makecell{0.2125\\(0.0064)} & \makecell{0.2393\\(0.0053)} & \makecell{\textbf{0.2033}\\(0.0093)} & 0.2906 & 0.3257 \\
\midrule
\multirow{3}{*}{\textbf{NASDAQ}}
& 1 & \makecell{0.1288\\(0.0125)} & \makecell{0.1190\\(0.0041)} & \makecell{0.2219\\(0.0075)} & \makecell{\textbf{0.1137}\\(0.0046)} & 0.2761 & 0.2766 \\
& 3 & \makecell{\textbf{0.1732}\\(0.0108)} & \makecell{0.1786\\(0.0070)} & \makecell{0.2268\\(0.0075)} & \makecell{0.1785\\(0.0030)} & 0.3213 & 0.3233 \\
& 7 & \makecell{\textbf{0.2552}\\(0.0064)} & \makecell{0.2587\\(0.0093)} & \makecell{0.3098\\(0.0048)} & \makecell{0.2567\\(0.0045)} & 0.4017 & 0.4169 \\
\bottomrule
\end{tabular}
}
\captionsetup{justification=raggedright, singlelinecheck=false}
\caption*{\scriptsize Results are based on the 2019 out-of-sample evaluation. Reported values denote the mean and standard deviation across 20 seeds where applicable.}
\end{table}

\end{document}